\documentclass[a4paper,fleqn]{cas-sc}

\usepackage[numbers]{natbib}

\usepackage{graphicx}%
\usepackage{multirow}%
\usepackage{amsmath,amssymb,amsfonts}%
\usepackage{amsthm}%
\usepackage{mathrsfs}%
\usepackage[title]{appendix}%
\usepackage{xcolor}%
\usepackage{textcomp}%
\usepackage{manyfoot}%
\usepackage{booktabs}%
\usepackage{algorithm}%
\usepackage{algorithmicx}%
\usepackage{algpseudocode}%
\usepackage{listings}%
\usepackage{physics}
\usepackage{optidef}
\usepackage{bm}

\def\tsc#1{\csdef{#1}{\textsc{\lowercase{#1}}\xspace}}
\tsc{WGM}
\tsc{QE}
\tsc{EP}
\tsc{PMS}
\tsc{BEC}
\tsc{DE}

\usepackage{color}
\usepackage[normalem]{ulem}
\newcommand{\Add}[1]{\textcolor{black}{#1}} 
\newcommand{\Del}[1]{\if0{#1}\fi} 

\begin{document}
\let\WriteBookmarks\relax
\def\floatpagepagefraction{1}
\def\textpagefraction{.001}
\shorttitle{Model Evaluation of a Transformable CubeSat for Nonholonomic Attitude Reorientation Using a Drop Tower}
\shortauthors{Yuki Kubo et~al.}

\title [mode = title]{Model Evaluation of a Transformable CubeSat for Nonholonomic Attitude Reorientation Using a Drop Tower}                      
\tnotemark[1,2]

\tnotetext[1]{This document is the result of the research
   project funded by the Grant for a strategic research group from the Advisory Committee for Space Engineering in Japan.}
\tnotetext[2]{The preprint version of this paper was published online at https://doi.org/10.48550/arXiv.2501.17173}

\author[1]{Yuki Kubo}[type=editor,
                        auid=000,bioid=1,
                        orcid=0000-0001-7710-9504]
\cormark[1]
\ead{kubo.yuki@harbor.kobe-u.ac.jp}
\ead[url]{https://researchmap.jp/yukikubo}

\credit{Conceptualization of this study, Methodology, Data curation, Project administration, Writing, Investigation}

\affiliation[1]{organization={Faculty of System Informatics, Kobe University},
                addressline={1-1 Rokkodai-cho, Nada-ku}, 
                city={Kobe},
                postcode={657-8501}, 
                state={Hyogo},
                country={Japan}}

\author[2]{Tsubasa Ando}
\ead{c5624137@aoyama.jp}
\credit{Data curation, Software, Investigation}

\author[2]{Hirona Kawahara}[]
\ead{c5625187@aoyama.jp}
\credit{Data curation, Software, Investigation}

\author[2]{Shu Miyata}
\ead{a5622118@aoyama.jp}
\credit{Investigation}

\affiliation[2]{organization={Department of Mechanical Engineering, Aoyama Gakuin University},
                addressline={5-10-1 Fuchinobe}, 
                postcode={252-5258}, 
                city={Chuo-ku, Sagamihara},
                state={Kanagawa},
                country={Japan}}

\author[2]{Naoya Uchiyama}
\ead{naoyauchikmn@gmail.com}
\credit{Investigation}


\author[3]{Kazutoshi Ito}
\ead{kazu-physics6.97@fuji.waseda.jp}
\credit{Investigation, Methodology, Software}

\affiliation[3]{organization={Department of Applied Mechanics and Aerospace Engineering, Waseda University},
                addressline={3-4-1 Okubo}, 
                city={Shinjuku-ku},
                postcode={169-8555}, 
                state={Tokyo}, 
                country={Japan}}

\author[2]{Yoshiki Sugawara}
\ead{sugawara@me.aoyama.ac.jp}
\credit{Supervision}

\cortext[cor1]{Corresponding author}


\begin{abstract}
This paper presents a design for a drop tower test to evaluate a numerical model for a structurally reconfigurable spacecraft with actuatable joints, referred to as a transformable spacecraft. A mock-up robot for a 3U-sized transformable spacecraft is designed to fit \Add{within the}\Del{in a} limited time and space \Add{constraints} of the microgravity environment available in the drop tower. The robot performs agile reorientation, referred to as nonholonomic attitude control, by actuating joints in a particular manner. To adapt to the very short duration of microgravity in the drop tower test, a successive joint actuation maneuver is optimized to maximize the amount of attitude reorientation within the time constraint. The robot records the angular velocity history of all four bodies, and the data is analyzed to evaluate the accuracy of the numerical model. We confirm that the constructed numerical model sufficiently replicates the robot's motion and show that the post-experiment model corrections further improve the accuracy of the numerical simulations. Finally, the difference between this drop tower test and the actual orbit demonstration is discussed to show the prospect.
\end{abstract}



\begin{keywords}
    Drop tower test \sep Reconfigurable spacecraft \sep Attitude reorientation \sep Nonholonomic system \sep Particle swarm optimization \sep CubeSat
\end{keywords}

\maketitle

\section{Introduction}\label{sec:intro}
Novel space exploration technologies have grown rapidly and widely in recent years. \Add{Future space robots (including spacecraft in a broad sense) will need to be, for example, more autonomous, more intelligent, more adaptive, more dexterous, and more economical.} A transformable spacecraft is one \Add{promising}\Del{such} architecture, enabling a spacecraft to reconfigure itself with actuatable joints. We can expect the transformable spacecraft to perform multi-objective operations with highly redundant actuators. 
The Transformer research group, organized by JAXA and universities in Japan, has proposed and studied a concept of such a highly redundant "Transformer" \citep{sugawara2020iac}. 
The other group, Tokyo Institute of Technology, launched HIBARI in 2021 \citep{watanabe2023flight,kobayashi2024lessons}. HIBARI had only four degrees of freedom actuators, but it carried out important and insightful orbit demonstrations with actuatable mechanisms. \par
One key technology for the transformable spacecraft is "nonholonomic attitude control," where the spacecraft can reorient its attitude by successive joint actuation\Del{(the details are described in Section \ref{subsec:nh})}. 
\Add{The study of nonholonomic systems has a long history. Comprehensive surveys can be found in textbooks such as that by Bloch \cite{bloch2003nonholonomic}.
As for the mechanics of free-flyers, one of the most widely known nonholonomic problems is the righting reflex of falling cats (or the falling-cat phenomenon). This is a motion in which a cat can always land in an upright attitude by twisting its body in the air, even when dropped in an upside-down attitude.
There have been many studies to elucidate the dynamics of this phenomenon \citep{muller1916notes}\citep{kane1969dynamical}.
Since the 1980s, when numerous space robots were proposed to construct space stations \citep{akin1983space}\citep{bronez1986requirements}, more studies have focused on the dynamics of free-flying robots.
Some researchers have proposed actively using the attitude reaction when a manipulator is actuated. Many attitude reorientation methods have been proposed that leverage the nonholonomy of the attitude motion
\citep{nakamura1993exploiting}\citep{mukherjee1999almost}\citep{yamada1997feedback}\citep{suzuki1996planning}\citep{cerven2001optimal}\citep{hokamoto2007feedback}\citep{kojima2011adaptive}\citep{lee2011applications}\citep{trovarelli2020attitude}.
}\par
Physical properties such as mass, position of the center of mass, and moment of inertia determine the direction and amount of the reorientation \Add{of the nonholonomic attitude control}. Therefore, understanding the fidelity and accuracy of the numerical model is important for the maneuver design. 
In general, a spacecraft is composed of many components and is not necessarily precisely modeled due to time and cost constraints. Moreover, some components, such as flexible harnesses and liquid propellants, cannot be modeled precisely. 
Thus, measuring such physical properties through ground tests and operations in orbit is essential. 
\par
\Add{Hardware-based experiments are essential to evaluate the modeling accuracy of the physical properties of robots.}
There are \Add{several}\Del{a few} examples of experiments to demonstrate the nonholonomic attitude control in hardware. 
\Add{Some researchers attempted to apply the righting reflex of falling cats to free-flying robots or spacecrafts \citep{yamafuji1992elucidation}\citep{li2018landing}\citep{garant2020design}.}
\Add{The}\Del{One of the} most active projects in recent years \Add{are}\Del{is} carried out by ETH Zurich, named "SpaceBok" \cite{rudin2021cat} and "SpaceHopper" \cite{spiridonov2024spacehopper}. Both are legged robots designed to walk and jump in a low-gravity space environment. After a jump on the surface, the robot changes its attitude by nonholonomic attitude control using its legs. To construct the reorientation control law, they adopted deep reinforcement learning in a simulation environment. The simulation parameters, such as masses, moment of inertia, and friction\Del{s} coefficients, cannot be precisely modeled in the simulator, and thus, they used the method of domain randomization \cite{tobin2017domain}. The policy was trained on many environments in parallel, each having a randomized set of simulation parameters with 20--50\% uncertainty. However, such uncertainty may deteriorate the capability of attitude control. In this situation, \Add{evaluating and updating}\Del{evaluation and update of} such physical parameters \Add{using}\Del{by} the obtained experimental data can contribute to enhancing attitude control capability. \par
This article describes a method of model evaluation using a drop tower facility. We introduce the experiment's design and analyze the obtained results. \Add{The motivation of the experiment is to estimate the modeling error and the performance of attitude reorientation. Moreover, by extrapolating the result obtained in the drop tower test, we can estimate the on-orbit reorientation performance assuming a certain 3U transformable CubeSat.}
Section \ref{sec:hardware} describes the facility of the drop tower, the design of a transformable robot, and a release mechanism. Section \ref{sec:maneuver} first introduces an overview of the nonholonomic attitude control and is followed by physical modeling and maneuver designs using the particle swarm optimization. Section \ref{sec:setup} explains setups of the experiment and is followed by results\Del{and discussions} in Section \ref{sec:result}\Del{ and \ref{sec:discussion}}. \Add{Finally, Section \ref{sec:discussion} discusses expected on-orbit performance and several differences between the ground and on-orbit experiment.}
The preprint version of this article can be found in the arXiv server \cite{kubo2025modelevaluationtransformablecubesat}.

\section{Design of hardware}\label{sec:hardware}
This section describes the experiment facility and the hardware design. 
\subsection{Facility of the drop tower}\label{subsec:droptower}
\begin{figure*}[h]
    \centering
    \includegraphics[width=0.9\textwidth]{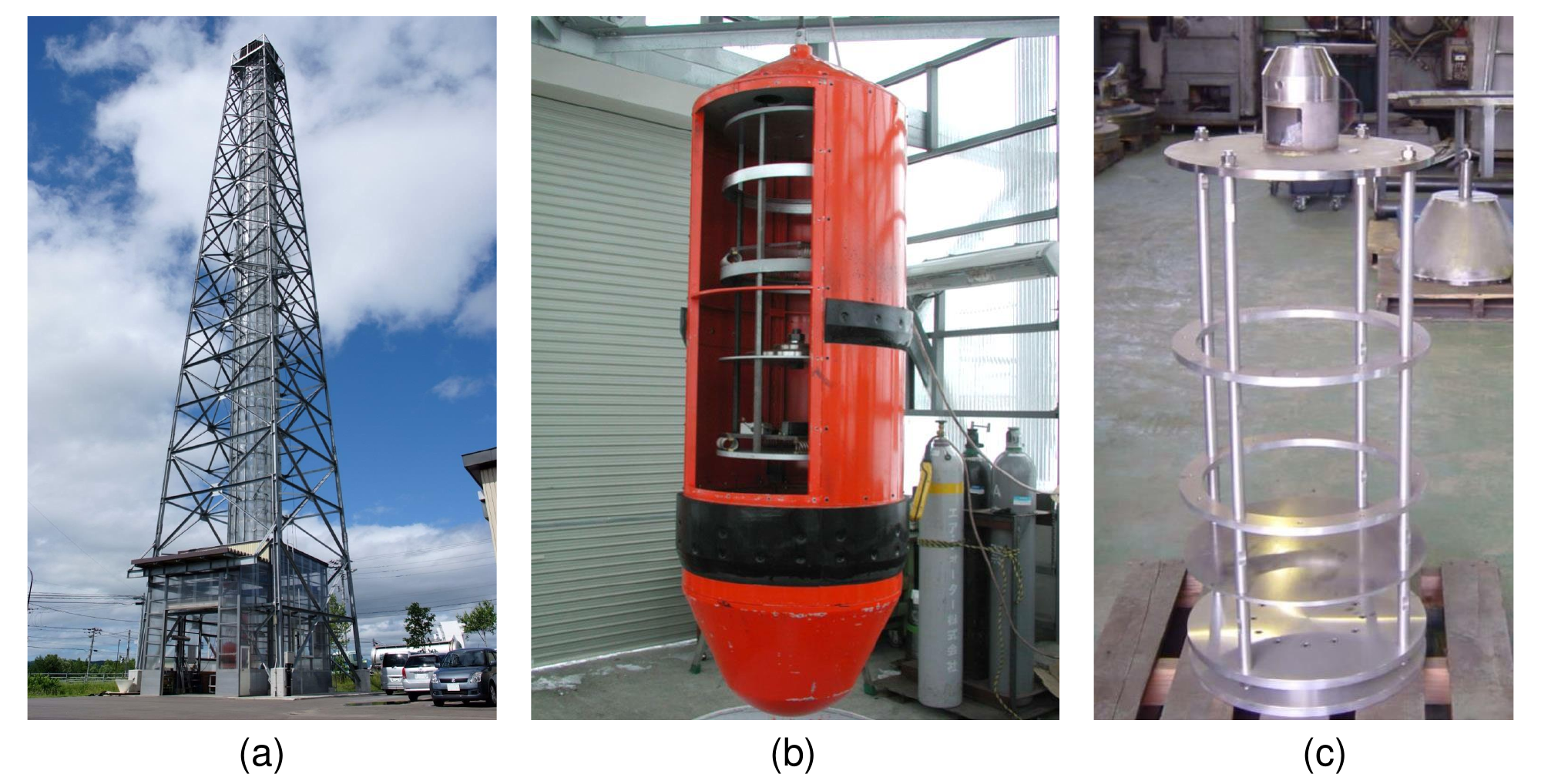}
    \caption{(a) COSMOTORRE, (b) an outer capsule, (c) an inner capsule}\label{fig:cosmotorre}
\end{figure*}

A drop tower is less costly than an orbit demonstration and a parabolic flight by an airplane. 
In addition, the test hardware can be refurbished on the ground during relatively long intervals \Add{between}\Del{of} capsule drops. 
For these reasons, we adopted the drop tower as a test facility for this experiment.
In particular, the drop tower facility COSMOTORRE in Hokkaido, Japan, was chosen for this experiment. The tower and the drop capsule are shown in Fig. \ref{fig:cosmotorre}, and their specification is overviewed in Table \ref{tab:cosmotorre}. Instruments for the experiment are equipped in the inner capsule. The inner capsule is installed inside the outer capsule, which reduces air drag during free fall. 

\begin{table}[h]
\caption{Specification overview of COSMOTORRE}\label{tab:cosmotorre}%
\begin{tabular}{@{}ll@{}}
\toprule
Drop height & 45 m \\
Micro-G duration & 2.5 sec  \\
Micro-G quality & $> 10^{-3}$ G \\
Payload size & $\phi 50 \times 105$ cm \\
Maximum payload mass & 400 kg \\
\bottomrule
\end{tabular}
\end{table}

\subsection{Design of the transformable robot}\label{subsec:robot}
\begin{figure*}[h]
    \centering
    \includegraphics[width=0.8\textwidth]{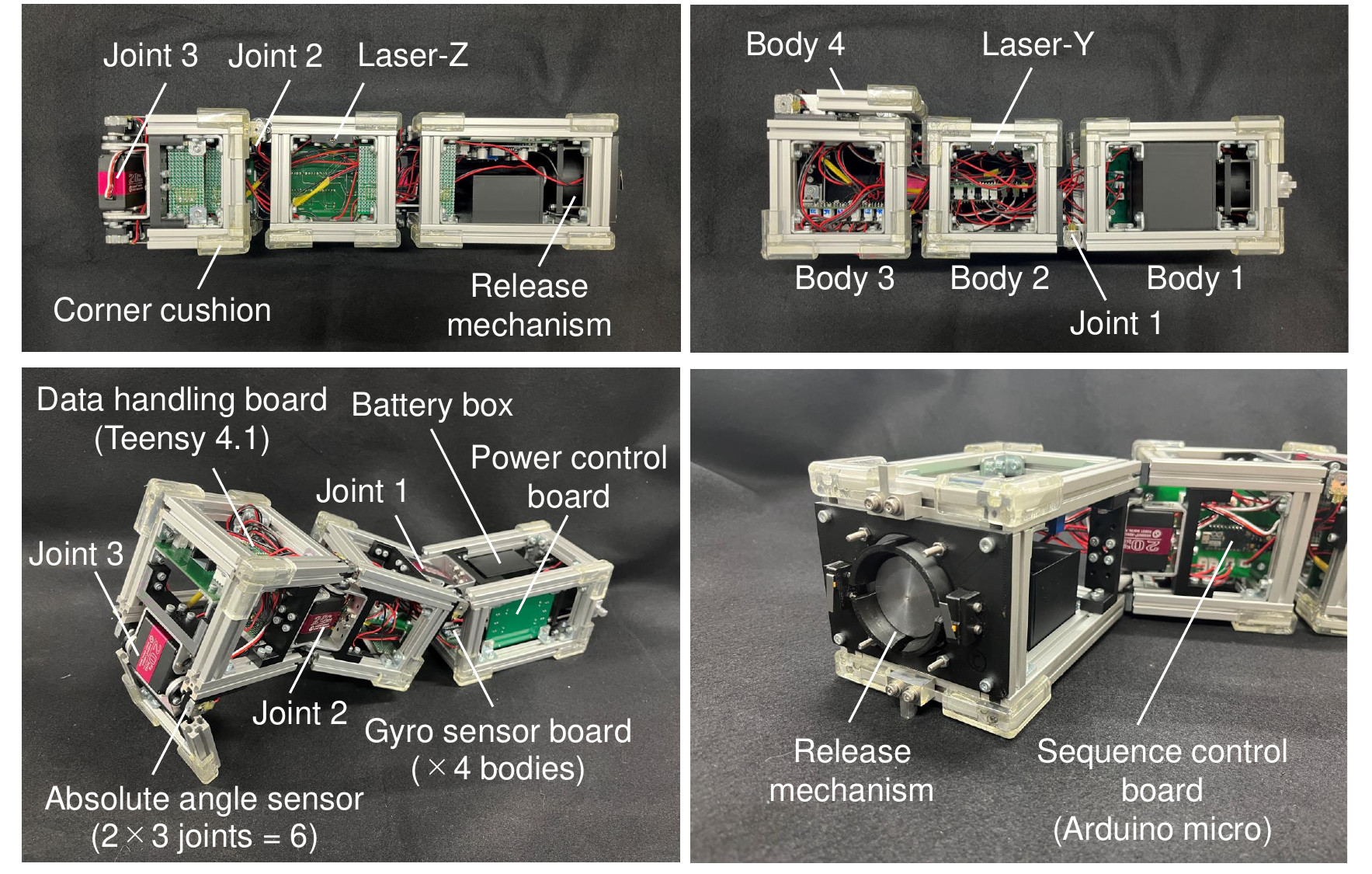} 
    \caption{Overview of the transformable robot}\label{fig:robot_ov}
\end{figure*}

\begin{figure*}[h]
    \centering
    \includegraphics[width=0.7\textwidth]{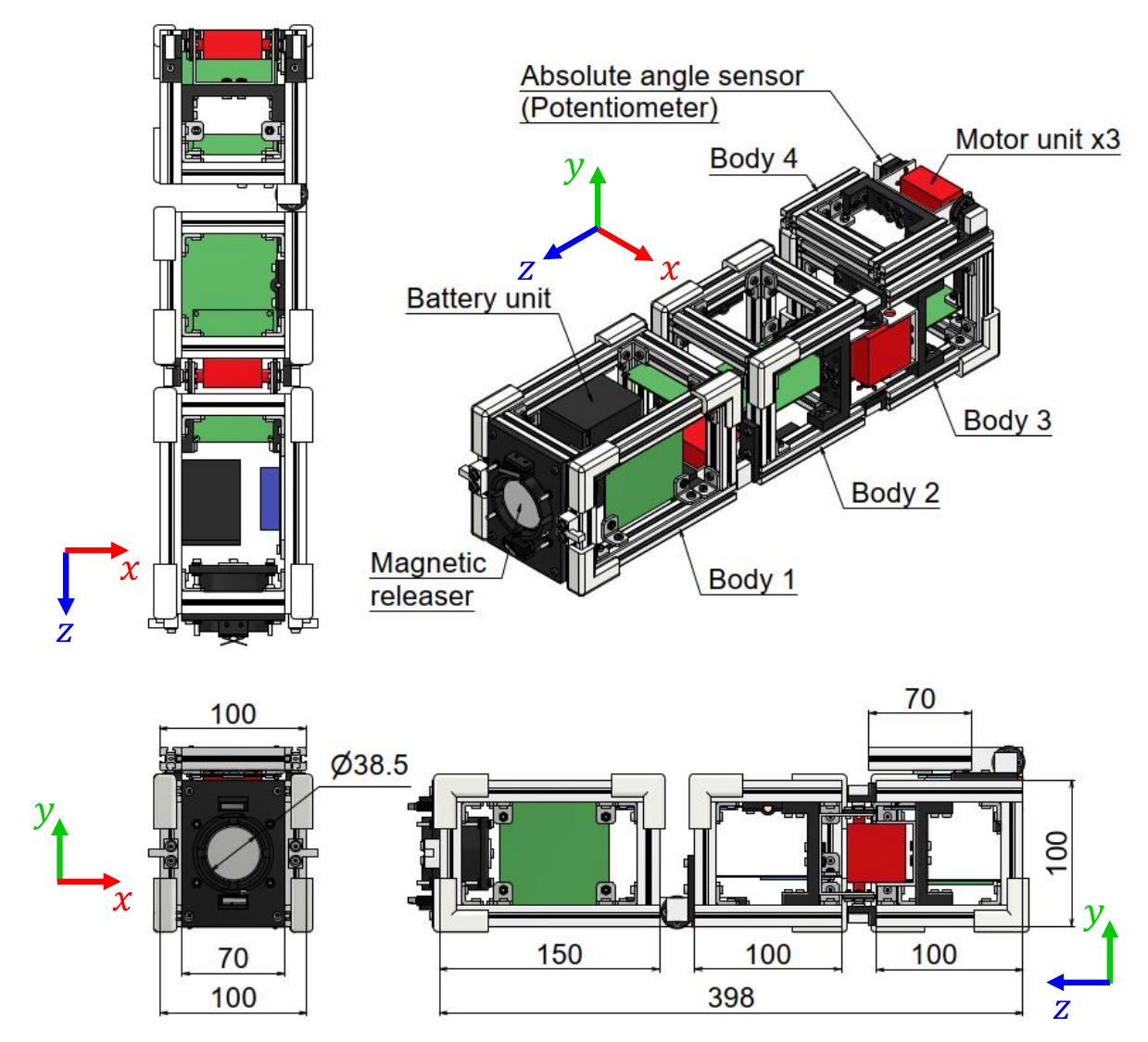} 
    \caption{Dimensions of the transformable robot}\label{fig:robot_dim}
\end{figure*}

The transformable robot in this experiment is designed to satisfy some requirements imposed by the drop tower facility:
\begin{itemize}
    \item To finish a set of maneuvers during the micro-G duration of the drop tower (about 2.5 sec)
    \item To fit in the inner capsule envelope ($\phi 50 \times 105$ cm) during the entire maneuvers
\end{itemize}
The overview and dimensions of the designed robot are shown in Fig. \ref{fig:robot_ov} and \ref{fig:robot_dim}. The robot is composed of four bodies with three actuatable joint units. Each joint unit comprises a servomotor, two absolute angle sensors (at both ends of the rotational axis of the motor), and bracket structures. The total mass is 2,847 g. The frame structure is composed of universal aluminum frames that can endure the landing impact at the end of the drop test. The robot's size is compatible with \Add{a} 3U CubeSat, whereas the length in \Add{the} $x$-direction is elongated due to the release mechanism. \par
The joint is actuatable in a range of 90 degrees, with 20 kgf$\cdot$cm torque. With this setup, the average angular velocity of the joint is 150-200 deg/s; thus, the possible total angular displacement during 2.5 sec is 375-500 deg. \par
MEMS gyroscope boards equipped with InvenSense MPU-6050 are mounted on all four bodies. Parameter specifications of the MPU-6050 are shown in Table \ref{tab:mpu}.
\par
\begin{table}[h] \caption{Specifications of InvenSense MPU-6050}\label{tab:mpu}%
    \begin{tabular}{@{}ll@{}}
    \toprule
    Parameters & Values \\
    \midrule
    Full-scale range & $\pm$ 500 deg/s\\
    Full-scale bit length & 16 bits \\
    Bandwidth of low pass filter & 5 Hz \\
    Maximum sampling rate & 1 kHz \\
    Sensitivity scale factor tolerance & $\pm$ 3\% \\
    Linear acceleration sensitivity & 0.1 deg/s/g \\
    \bottomrule
    \end{tabular}
\end{table}
Two lasers are mounted orthogonally on body 2. The laser \Add{ray}\Del{lay} is projected on graph paper attached to the side walls of the test space and captured by cameras, which allows us to estimate the release velocity of the robot. \par
The robot has two microprocessors. \Add{The} Arduino micro\Add{,} mounted on the sequence control board\Add{,} controls the entire maneuver sequence of the robot. On the other hand, Teensy 4.1, mounted on the data handling board, specializes in data acquisition from MEMS gyroscopes and absolute angle sensors and data recording on an SD card. This separation of the roles enables the robot to achieve a sampling rate of about 280 Hz, which is sufficiently fast in this experiment.

\subsection{Design of the release mechanism}\label{subsec:release}
\begin{figure*}[h]
    \centering
    \includegraphics[width=0.9\textwidth]{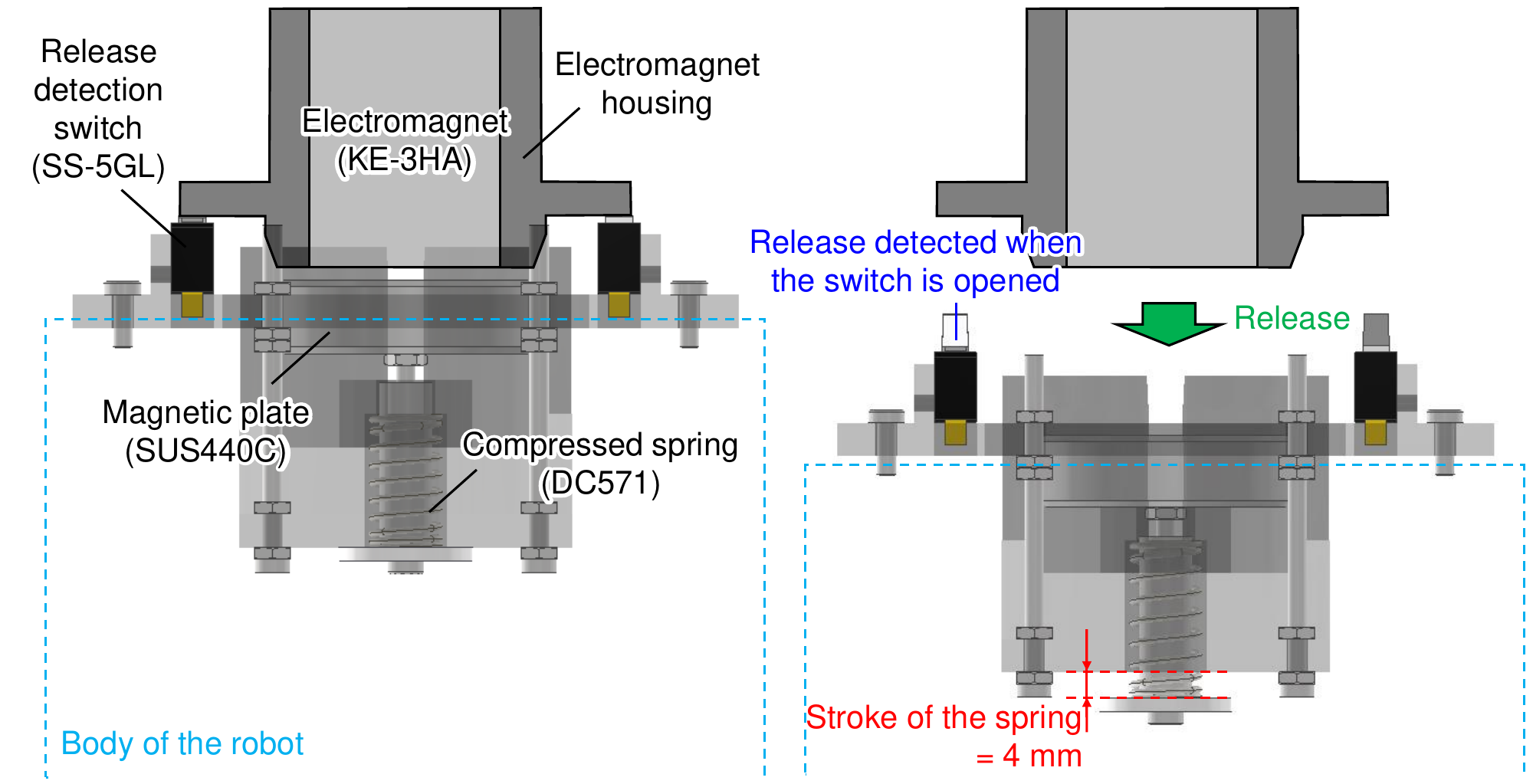} 
    \caption{Overview of the magnetic releaser}\label{fig:mag_rel}
\end{figure*}

The release mechanism is designed to satisfy the following requirements:
\begin{enumerate}
    \item To stably hold the robot until the beginning of the microgravity experiment
    \item To provide detaching force to overcome residual magnetic force under micro-gravity
    \item To push the robot with the proper vertical velocity at \Add{release}\Del{releasing} 
    \item To suppress transversal releasing velocity to prevent the robot from touching the frame structure during the entire maneuver
\end{enumerate}
Figure \ref{fig:mag_rel} shows the overview of the designed release mechanism. \par
\begin{figure}[h]
    \centering
    \includegraphics[width=0.35\textwidth]{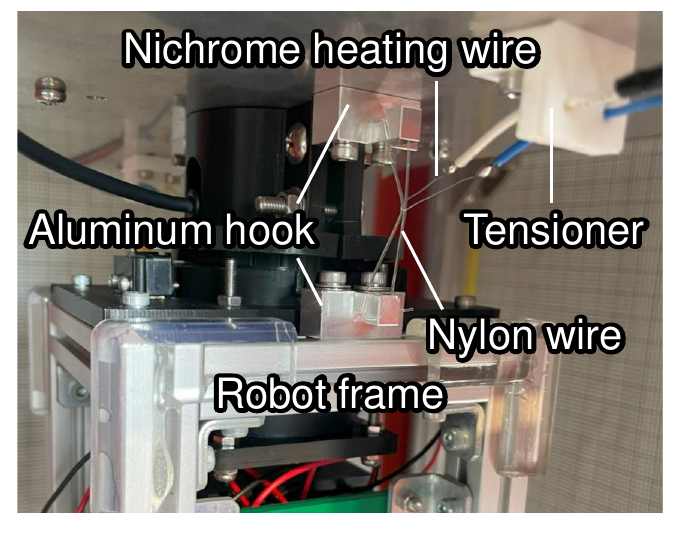} 
    \caption{Holding mechanism with nylon wire}\label{fig:nylon}
\end{figure}

The robot is attached to the electromagnet unit with the magnetic plate of SUS440C. The electromagnet, Kanetec KE-3HA, provides a magnetic force of 200 N (20.4 kgf) at a maximum, which can sufficiently hold the robot weighing about 3 kg. However, an extra holding mechanism with nylon wires shown in Fig. \ref{fig:nylon} is attached because higher acceleration is added when a crane lifts the capsule. After \Add{the} crane lifting is completed, the nylon wire is tied firmly between hooks and burned by a nichrome electric heating wire. Tensile force on the nylon wires is applied by sliding the position of wire tensioners, which enables tight holding and secure cutting. Requirement 1, shown above, is satisfied with this electromagnet and nylon wire holding mechanism. \par
The magnetic plate is connected to the shaft with a spring and fits in the 3D-printed housing structure. The spring is compressed when the robot is attached, and thus, it provides \Add{a} detaching force to overcome residual magnetic force under micro-gravity. The residual magnetic force is estimated as 5\% of the maximum magnetic force, hence 10 N, according to the datasheet. The designed spring mechanism provides 11.1 N, which can give sufficient detaching force under microgravity. In addition, the stroke of the spring is set to 4 mm, which can give sufficient separation distance (1 mm is enough according to the datasheet). Requirement 2 is satisfied with this mechanism. \par
Two release detection switches (OMRON SS-5GL) are equipped beside the magnetic plate. The switches are closed when the robot is attached to the electromagnet and opened when released. The robot starts a preset sequence after detecting the \Add{switch}\Del{switching}. In addition, the switches slightly push off the robots at the release and provide proper vertical velocity. If the vertical velocity is small, the robot must wait \Add{for a} very long \Add{time} until the beginning of movement to prevent itself from colliding with the magnetic releaser. On the other hand, if the vertical velocity is too large, the robot collides with the bottom of the test space before the end of the attitude maneuver. Thus, the proper vertical velocity range is estimated as $20<v_0<100$ mm/s. \par
\begin{figure}[h]
    \centering
    \includegraphics[width=0.45\textwidth]{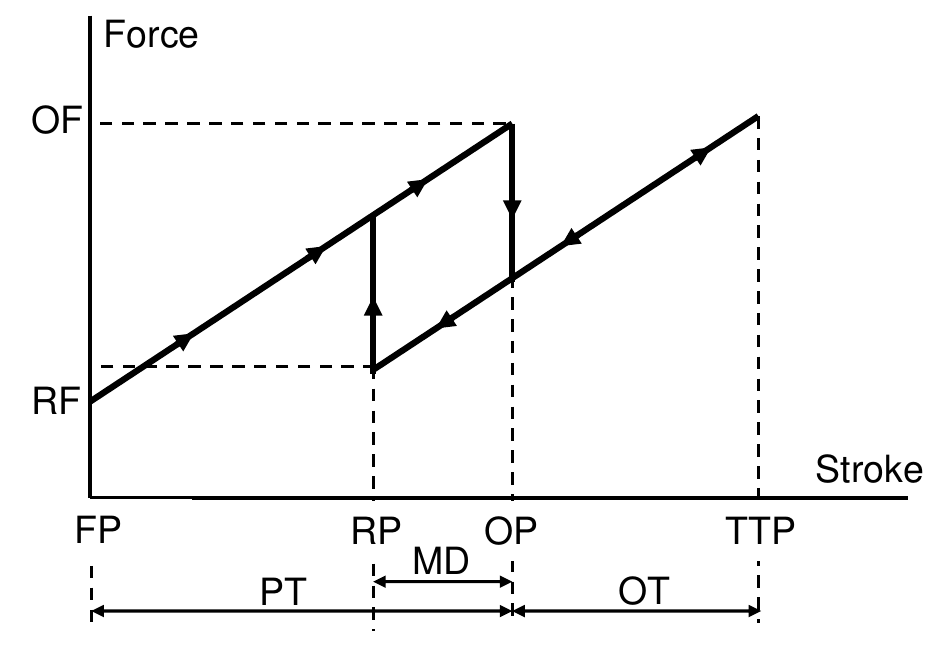} 
    \caption{Force-stroke plot for release detection switches}\label{fig:sw_en}
\end{figure}

Figure \ref{fig:sw_en} illustrates the relationship between pushing force and stroke of a switch lever. Moreover, the figure shows relationships of some operational characteristics: OF (operation force), RF (releasing force), PT (pre-travel), OT (over-travel), MD (movement differential), FP (free position), OP (operating position), RP (releasing position), and TTP (total travel position). The properties of SS-5GL are shown in Table \ref{tab:ss5gl}. From these properties, we can estimate the kinetic energy $K$ provided by the switches by calculating the area enclosed by the thick lines in the graph of Fig. \ref{fig:sw_en}. It is estimated as $2.35<K<3.12$ mJ with two switches, corresponding to a vertical releasing velocity of $40.6<v_0<46.8$ mm/s. 
The actual vertical velocity observed in the test was distributed in the $30<v_0<50$ mm/s because other factors, such as contact friction force between electromagnet housings, are complicated. However, the range is always within the required level, and thus, we concluded that requirement 3 is satisfied with this design. \par
\begin{figure}[h]
    \centering
    \includegraphics[width=0.3\textwidth]{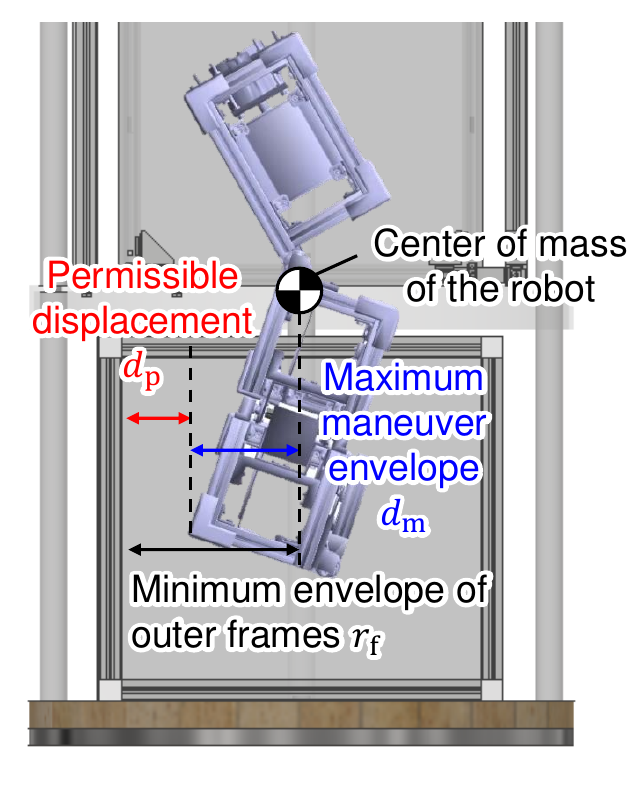} 
    \caption{Permissible transversal displacement of release}\label{fig:per_dis}
\end{figure}

\begin{table}[h] \caption{Operating properties of OMRON SS-5GL}\label{tab:ss5gl}%
    \begin{tabular}{@{}ll@{}}
    \toprule
    OF max. & 0.49 N \\
    RF min. & 0.06 N \\
    OT min. & 1.2 mm \\
    MD max. & 0.8 mm \\
    FP max. & 13.6 mm \\
    OP      & 8.8$\pm$0.8 mm \\
    \bottomrule
    \end{tabular}
\end{table}
Figure \ref{fig:per_dis} shows the relationship between the maximum maneuver envelope and permissible displacement during the maneuver. The release mechanism is required to release the robot as the transversal velocity is under the permissible level to prevent the robot from colliding with walls and pillars. In this experiment, the radius of the minimum envelope is 165 mm, whereas the maximum envelope of the robot during an entire maneuver is 100 mm. Thus the permissible transversal releasing velocity $v_\mathrm{p}$ during is calculated as:
\begin{equation}\label{eq:per_dis}
    \begin{split}
        v_\mathrm{p} &= 
        \frac{r_\mathrm{f} - d_\mathrm{m}}{t_\mathrm{maneuver}} \\
        &= \frac{165 \mathrm{mm} - 100 \mathrm{mm}}{2.5 \mathrm{s}} = 26 \mathrm{mm/s} 
    \end{split} 
\end{equation}
Throughout the preparation test, we confirmed that the designed releasing mechanism has satisfied this requirement. In the actual test shown in Section \ref{sec:result}, the transversal velocity was about 7 mm/s, and it is confirmed that there were no collisions with walls and pillars.

\section{Maneuver design}\label{sec:maneuver}
This section describes the design of the reorientation maneuver adopted in the experiment. First, we \Add{provide an} overview \Add{of} the principles of attitude reorientation and a fundamental maneuver in Section \ref{subsec:nh}. 
Section \ref{subsec:model} denotes the physical modeling of the robot. 

\subsection{Nonholonomic attitude reorientation}\label{subsec:nh}
\begin{figure*}[h]
    \centering
    \includegraphics[width=0.9\textwidth]{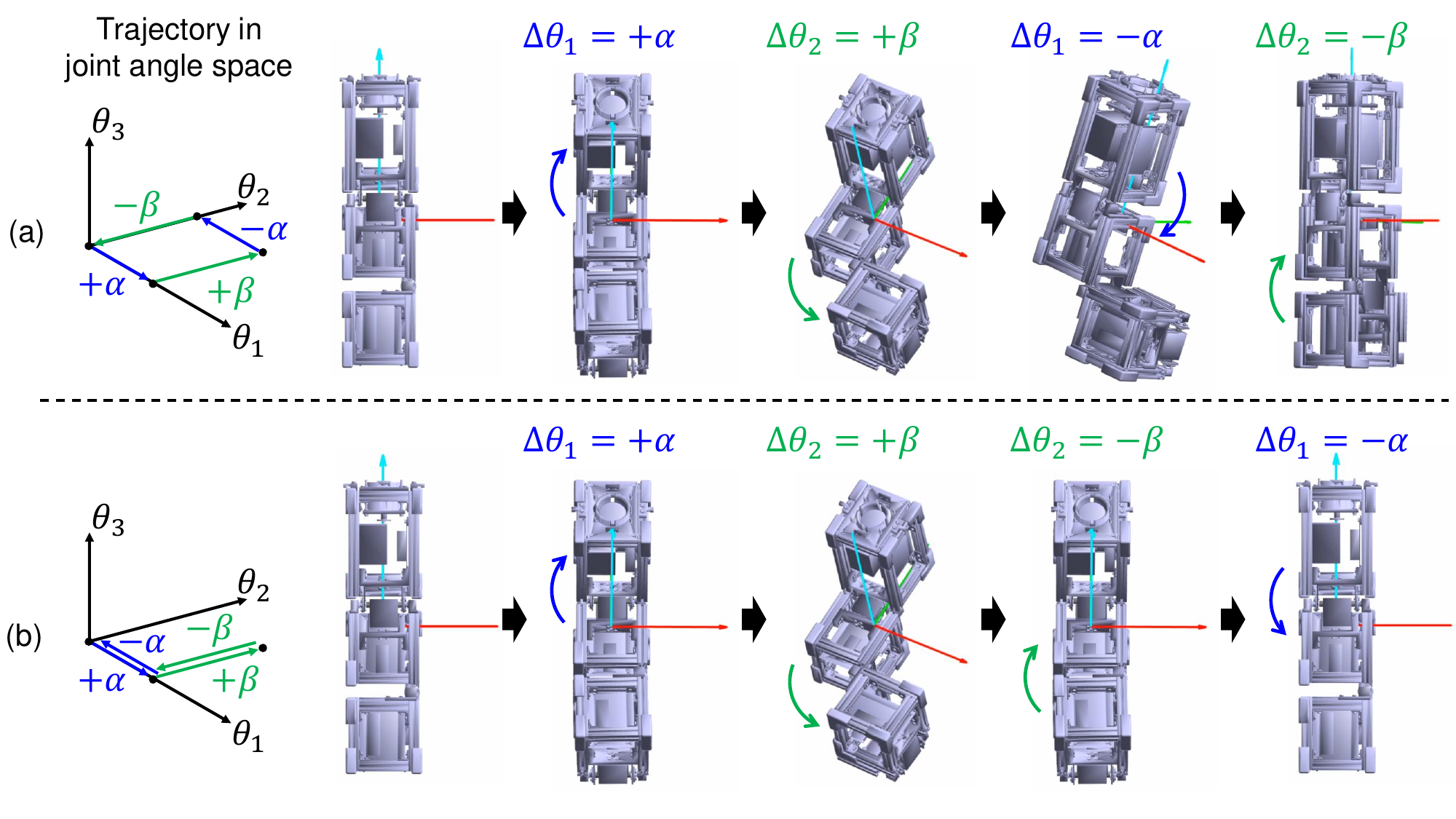} 
    \caption{Two examples of joint actuation (a): with reorientation, (b): without reorientation}\label{fig:ex_para}
\end{figure*}

The dynamics of free-floating \Add{multibodies}\Del{multibody} have been intensively investigated. It is known that the free-flying multibody can reorient itself by a proper sequence of joint actuation, even if the total angular momentum has always been kept at zero. Historically, it is known by \Add{the} righting reflex of falling cats \citep{muller1916notes,mcdonald1955does,kane1969dynamical}, and intensively studied for space robots with robotic manipulators \citep{murray1993nonholonomic,dubowsky1993kinematics,nakamura1993exploiting,papadopoulos1993nonholonomic,mukherjee1999almost}. 
The angular momentum conservation law imposed on the free-flying robot is nonholonomic, which means the robot's final attitude depends on the procedure of joint actuation. Therefore, the robot can arbitrarily reorient its attitude by designing proper joint actuation maneuvers.
In this experiment, we adopt a maneuver similar to the one proposed in previous research, such as \cite{ohashi_aas2018, kubo2022nonholonomic}. Two examples of joint actuation are shown in Fig. \ref{fig:ex_para}. The sequence shown \Add{on}\Del{in} the upper side is composed of four consecutive joint actuations, where the first stroke and third strokes are actuations of the same joint but in the opposite direction ($\Delta\theta_1=+\alpha$ and $-\alpha$), and so do the second and fourth strokes ($\Delta\theta_2=+\beta$ and $-\beta$). The attitude is reoriented due to the nonholonomy of the system, and thus it is referred to as {\it nonholonomic attitude reorientation}. The one in the lower is composed of four consecutive joint actuation\Add{s}, but the first and fourth strokes are the same pattern as the second and third strokes. This sequence exactly cancels the intermediate attitude difference, and therefore, attitude is not reoriented after the four strokes. These two examples show that the final attitude depends on the intermediate joint actuation sequence. \par
The corresponding trajectory in a joint angle space is shown \Add{on}\Del{in} the left side of Fig. \ref{fig:ex_para}. 
A joint angle space is a multi-dimensional space where each point corresponds to one set of joint states. In these examples, the maneuvers are expressed as a consecutive rectilinear path. If the path draws a closed polygon as in the upper case in Fig. \ref{fig:ex_para}, the joint state of a robot returns to the initial state, whereas \Add{the} attitude is changed from the initial. This closed-loop actuation is adopted in this experiment because it clearly shows the effect of reorientation due to the nonholonomy of the system.

\subsection{Physical modeling of the robot}\label{subsec:model}
This section describes how to model physical properties \Add{of the robot. The robot is modeled as a multi-rigid-body system, and thus, the required physical properties to describe the robot are the mass, the position of}\Del{such as} center of mass, \Add{the} moment of inertia, \Add{directions of joints,} and \Add{the}\Del{joint} positions of joints \Add{relative to the center of mass of adjacent bodies}. \Add{The numerical model is}\Del{These properties are} necessary to design optimal reorientation maneuvers in Section \ref{subsec:opt}.\par
\begin{figure}[h]
    \centering
    \includegraphics[width=0.45\textwidth]{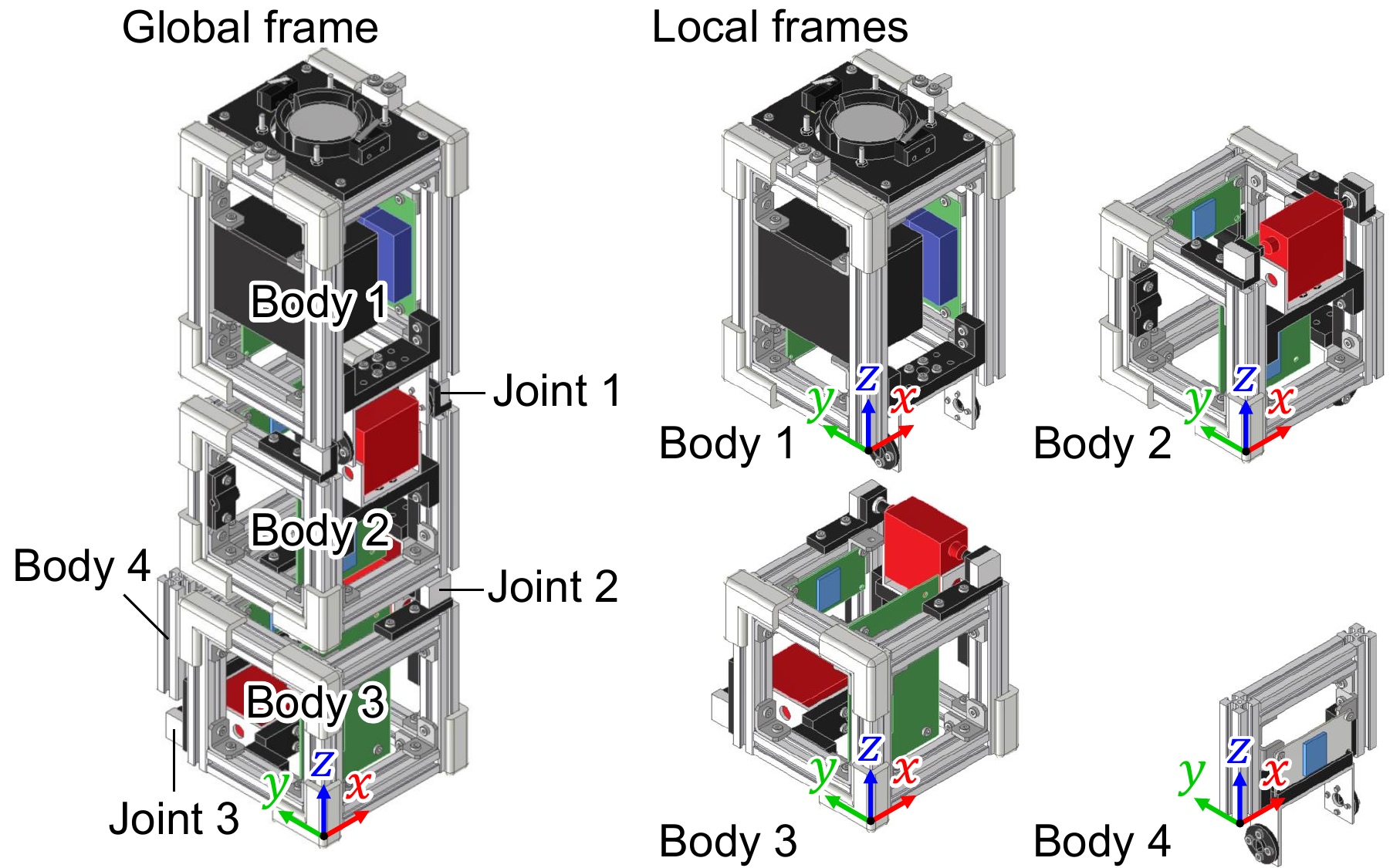} 
    \caption{\Add{Definitions of global body frame and local body frames}}\label{fig:bf}
\end{figure}

We modeled the robot based on the following assumptions:
\begin{itemize}
    \item Actual measurement value by an electric scale is used for a mass of each component 
    \item Center of mass and moment of inertia of each component are calculated assuming the component has homogeneous density
    \item Movable components such as wires are not included in the model 
\end{itemize}
Figure \ref{fig:bf} shows the definitions of global body frames and local body frames, indicating the origins and orientations of the frames. The origins are attached to the corner point of the aluminum frame.
Table \ref{tab:jp} shows the positions of the rotational axes of joints in the global frame \Add{and the local frame}. \Add{The positions in the global frame are the values when all the joint angles are set to be zero. In addition, the local position of the $i$-th joint is expressed in the local frame of the $(i+1)$-th body.} Note that the axes are aligned parallel to $x$ or $y$ in the global frame. Therefore, for example, the $x$ position of joint 1 is not indicated since the axis is parallel to the global $x$ axis.
Table \ref{tab:cg} shows the positions of the center of mass of each body in the local body frame. 
Table \ref{tab:il} is the moment of inertia of each body around the local center of mass in the local body frame.
\Add{All these physical properties include modeling errors due to the assumptions described above. In particular, the center of mass position and the moment of inertia have relatively large modeling errors because they cannot be directly measured from the actual hardware. In this article, we attempt to compensate for the modeling error using the experimental data in Section \ref{sec:result}.}

\begin{table}[h] \caption{Positions of rotational axes of joints in the global body frame (unit: mm)}\label{tab:jp}%
    \begin{tabular}{@{}lllllll@{}}
    \toprule
     & $x$ \Add{global} & $y$ \Add{global} & $z$ \Add{global} 
     & \Add{$x$ local} & \Add{$y$ local} & \Add{$z$ local} \\
    \midrule
    Joint 1 & - & \Add{10.0}\Del{9.9} & 236.0 & \Add{-} & \Add{10.0} & \Add{112.0}\\
    Joint 2 & \Add{90.0}\Del{90.1} & - & 112.0 & \Add{90.0} & \Add{-} & \Add{112.0}\\
    Joint 3 & - & 112.0 & 10.0 & \Add{-} & \Add{5.0} & \Add{-25.0}\\
    \bottomrule
    \end{tabular}
\end{table}

\begin{table}[h] \caption{Positions of the center of mass of each body in the local body frame (unit: mm)}\label{tab:cg}%
    \begin{tabular}{@{}llll@{}}
    \toprule
     & $x$ & $y$ & $z$ \\
    \midrule
    Body 1 & 46.58 & 50.93 & 84.18 \\
    Body 2 & 47.86 & 46.68 & 54.36 \\
    Body 3 & 52.76 & 53.17 & 48.77 \\
    Body 4 & 50.50 &  6.59 & 31.00 \\
    \bottomrule
    \end{tabular}
\end{table}

\begin{table}[h] \caption{Moment of inertia of each body around the local center of mass in the local body frame (unit: $\mathrm{g}\cdot \mathrm{m}^2$)}\label{tab:il}%
    \begin{tabular}{@{}lllllll@{}}
    \toprule
     & $I_{xx}$ & $I_{xy}$ & $I_{xz}$ & $I_{yy}$ & $I_{yz}$ & $I_{zz}$\\
    \midrule
    Body 1 & 3.97 & -0.01 & -0.08 & 4.10 &  0.09 & 2.27 \\
    Body 2 & 1.72 &  0.00 & -0.07 & 1.67 &  0.08 & 1.56 \\
    Body 3 & 2.00 & -0.01 & -0.09 & 1.94 &  0.09 & 1.80 \\
    Body 4 & 0.12 &  0.00 &  0.00 & 0.28 & -0.01 & 0.17 \\
    \bottomrule
    \end{tabular}
\end{table}

\subsection{Design of optimal reorientation maneuver}\label{subsec:opt}
We derive optimal reorientation maneuvers based on the model described in Section \ref{subsec:model}. The optimization is essential due to the following reasons:
\begin{itemize}
    \item The robot must complete the entire reorientation sequence within a very short duration of micro-gravity (around 2.5 seconds)
    \item It is necessary to achieve as large joint angle displacement as possible to observe the nonholonomic reorientation effect clearly
\end{itemize}
\Add{First, we set a constraint that only one joint can be moved in one stroke because the actuations of motors used in this experiment cannot be synchronized due to the hardware limitation. To identify a joint actuation sequence that induces a large attitude difference, we optimized joint angle displacements and the joint movement order separately. When optimizing the joint angle displacements, gradient-based optimization is difficult to apply because the objective function may be non-smooth in the design space. Therefore,}\Del{In this experiment,} we adopted the particle swarm optimization (PSO) \Add{to identify the joint angle displacements}. PSO is a bio-inspired \Add{meta}heuristic optimization method proposed in \cite{kennedy1995particle}. The formulation is simple and requires no gradient information for the objective function. Moreover, it is easily extensible to the robot with more actuatable joints and, thus, directly applicable in future missions.\par
\Add{In PSO, $N$ particles are prepared in the design space, and their initial positions and velocities are set randomly in the space. Then, the particles are updated according to the following rule \cite{wang2018particle}:}
\Add{
\begin{equation}
    \begin{split}
        x_{k+1}^i 
        \leftarrow& x_k^i + v_{k+1}^i
        ,\\
        v_{k+1}^i 
        \leftarrow& w v_{k}^i + c_1 r_1 \qty(x_{k}^{pi}-x_k^i) 
        + c_2 r_2\qty(x_k^g-x_k^i)
    \end{split}
\end{equation}
}
\Add{
where $x_k^i$ and $v_k^i$ are the position and the velocity of the $i$-th particle in the design space at the $k$-th iteration, $w$ is the inertia weight, $c_1$ and $c_2$ are the learning factors. Large $c_1$ imposes relatively high importance on the experience of each particle, whereas large $c_2$ imposes relatively high importance on the experience of all particles. $r_1$ and $r_2$ are random numbers sampled uniformly from 0.0 to 1.0, and $x_k^{pi}$ and $x_k^g$ are so far the best positions of the $i$-th particle and all the particles at the $k$-th iteration. Here, the position with the smallest objective function is selected as the best one. The updates are repeated until the number of iterations $k$ reaches $n$. To achieve efficient exploration of particles in PSO, coefficients $w$, $c_1$, $c_2$ are linearly changed according to the number of iterations $k$ in the following rule \cite{wang2018particle}:
\begin{equation}
    \begin{split}
        w =& 0.9-\frac{k}{2n}
        , \\
        c_1 =& 3.5-\frac{k}{n}
        ,\\
        c_2 =& 0.5+\frac{3k}{n}
    \end{split}
\end{equation}
Note that we determined the constants in these equations through trial and error. In the above rules, the experience of each particle is given importance at the beginning of the update, and experiences of all particles are given more importance as the number of iterations increases. In addition, to balance the computational cost and the convergence speed of PSO, we set the number of particles to $N=48$, and the number of iterations to $n=50$.}\par
\begin{figure}[h]
    \centering
    \includegraphics[width=0.2\textwidth]{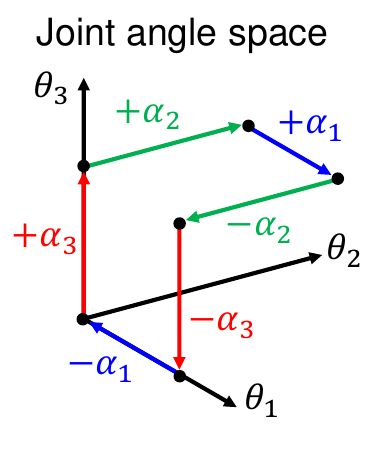} 
    \caption{An example of joint actuation sequence (the case of $(i,j,k)=(3,2,1)$, $+\alpha_i\rightarrow +\alpha_j\rightarrow +\alpha_k\rightarrow -\alpha_j\rightarrow -\alpha_i\rightarrow -\alpha_k$) }\label{fig:pso_ex}
\end{figure}
\begin{figure}[h]
    \centering
    \includegraphics[width=0.45\textwidth]{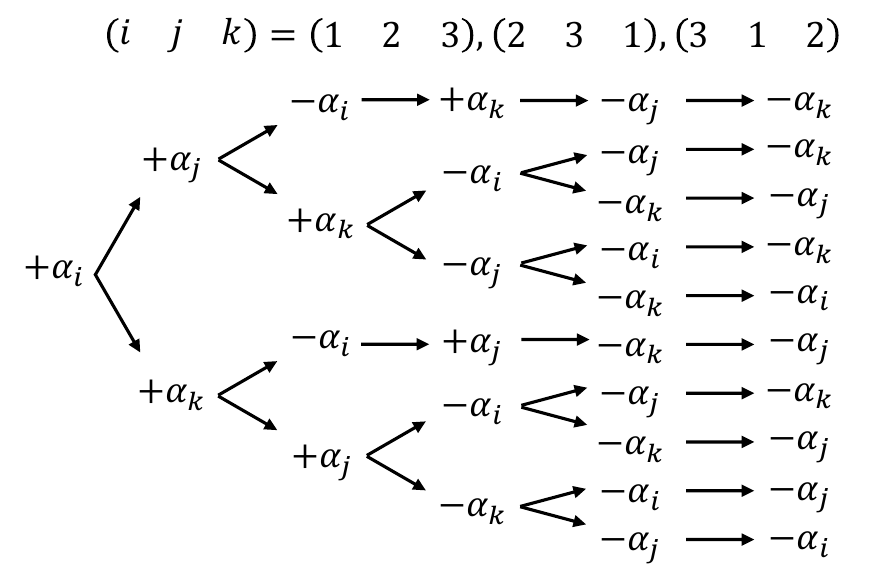} 
    \caption{All sequence pattern for a certain set of $(i,j,k)$}\label{fig:comb}
\end{figure}

The joint actuation sequence is expressed as a closed loop in a joint angle space as described in Section \ref{subsec:nh}. A schematic of an example of joint actuation sequence is shown in Fig. \ref{fig:pso_ex}. A point in the joint angle space corresponds to a set of specific joint angles. Angles of joints 1, 2, and 3 are denoted as $\theta_1,\theta_2,\theta_3$\Add{,} whereas input joint angle displacements are denoted as $\alpha_1,\alpha_2,\alpha_3>0$\Add{,} respectively. Specifically, Fig. \ref{fig:pso_ex} shows the case of $(i,j,k)=(3,2,1)$, $+\alpha_i\rightarrow +\alpha_j\rightarrow +\alpha_k\rightarrow -\alpha_j\rightarrow -\alpha_i\rightarrow -\alpha_k$. For a certain set of $(\alpha_1,\alpha_2,\alpha_3)$, the total number of combinations of joint actuation sequence is 30, as shown in Fig. \ref{fig:comb}. Note that $-\alpha_i$ is not chosen just after $+\alpha_i$ because it does not induce any attitude difference. 
\Add{In this experiment, we define the joint angle displacements $(\alpha_1, \alpha_2, \alpha_3)$ as the design parameters $x$. Then, we alternately perform optimizing the joint angles $(\alpha_1, \alpha_2, \alpha_3)$ via PSO and the joint movement sequence for a given set of them heuristically.}
\par
\begin{figure*}[h]
    \centering
    \includegraphics[width=0.9\textwidth]{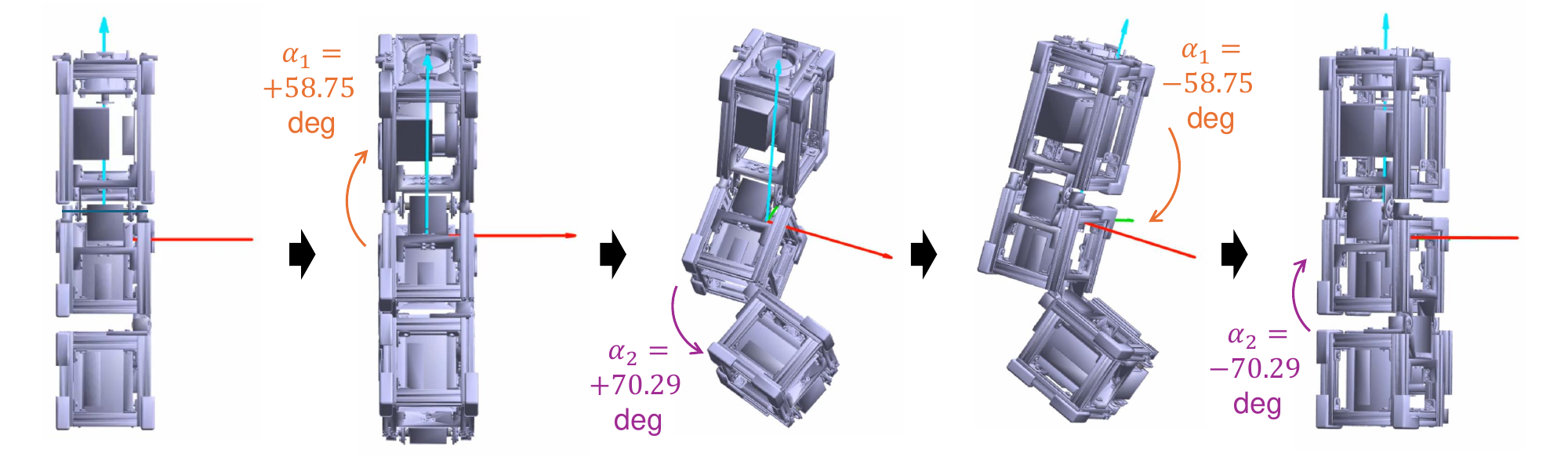} 
    \caption{The maneuver generated by the PSO}\label{fig:opt_man}
\end{figure*}

The optimization problem is defined as Eq. \eqref{eq:optdef}. The objective function $J$ is evaluated by the magnitude of attitude difference $\Delta\phi$ and the penalty on maneuver time overrun $P \qty(\max\qty[T-T_\mathrm{max},0])^2$. \Add{The penalty increases when $T>T_\mathrm{max}$, whereas it remains zero when $T\le T_\mathrm{max}$.} If $P$ is large, the maneuver time overrun is avoided more strictly. In this experiment, $P=10^{11}$ is chosen as a sufficiently large value. $T_\mathrm{max}$ is set to 1.5 sec, assuming a 1.0 sec margin should be included in the total 2.5 sec duration.
\begin{mini}
    {\alpha_1, \alpha_2, \alpha_3}          
    {J=\min_{C}\qty(-\Delta \phi + P \qty(\max\qty[T-T_\mathrm{max},0])^2)} 
    {\label{eq:optdef}}
    {} 
    \addConstraint{0\leq\theta_{1,2,3}}{\leq90\ \mathrm{deg}} 
    \addConstraint{\qty(\theta_1,\theta_2,\theta_3)}{=\qty(0,0,0)\ \mathrm{deg}\ \mathrm{at}\ t=0} 
\end{mini}
Applying PSO to the free-floating robot defined in Section \ref{subsec:model}, we obtained the optimized maneuver as shown in Fig. \ref{fig:opt_man}. Optimized joint angle displacements are $(\alpha_1,\alpha_2,\alpha_3)=(58.75,\ 70.29,\ 0.00)$ degrees. The average angular velocity of the joint is set to be 170 deg/s, resulting in 1.50 sec of the entire maneuver time. The resulting maneuver is composed only of 4 strokes because $\alpha_3$ is degenerated into 0 degrees through the optimization. This is because the moment of inertia of body 4 is smaller than \Add{the} others by one order\Add{,} as in Table \ref{tab:il}, and thus, the actuation of joint 3 is omitted in the limited maneuver time. The total attitude difference is 26.11 degrees in the $z$ axis. The resulting maneuver is tested in the drop tower test as shown in Section \ref{sec:result}. In the actual test, the values of the joint angle displacements commanded to motors are rounded as $(\alpha_1,\alpha_2)=(59, 70)$ to simplify the operations of the microprocessor.

\subsection{Design of experiment sequence}\label{subsec:seq}
Finally, we construct an entire sequence of the experiment with the resulting maneuver in Section \ref{subsec:opt}. Table \ref{tab:seq} shows the designed experiment sequence. \par
First, the capsule is dropped by the ground command. At almost the same time, the robot is released by the electromagnet controller that detects the microgravity. Soon, the release detection switch is opened, and the robot starts the movement sequence. These events occur almost simultaneously; thus, we ignore the delay between these events. The robot pauses to prevent touching the release mechanism during the first joint actuation. As described in Section \ref{subsec:release}, the worst (slowest) vertical releasing velocity is estimated as about $30$ mm/s. Thus, the pausing time is set to be $0.7$ seconds to guarantee the robot does not collide with the releaser, even in the worst case. \par
After the \Add{pause}\Del{pausing}, the robot starts the maneuver designed in Section \ref{subsec:opt}. The interval between the actuation\Add{s} is calculated as the joints move at 170 deg/s on average. After the four strokes, the robot pauses the joint actuation and continues MEMS gyroscope and joint angle data recording until the \Add{capsule decelerates}\Del{deceleration of the capsule}. 
\begin{table*}[h] \caption{Designed experiment sequence}\label{tab:seq}%
    \begin{tabular*}{\textwidth}{@{}ccl@{}}
    \toprule
    Sequence & Time [sec] & Events \\
    \midrule
    0 & 0.00 & Capsule dropped, electromagnet released, and release detected \\
    1 & 0.70 & Joint 1 commanded to be actuated ($+59$ degrees) \\
    2 & 1.05 & Joint 2 commanded to be actuated ($+70$ degrees) \\
    3 & 1.47 & Joint 1 commanded to be actuated ($-59$ degrees) \\
    4 & 1.82 & Joint 2 commanded to be actuated ($-70$ degrees) \\
    5 & 2.23 & Maneuver finished, and data recording maintained \\
    6 & 2.50 & Drop capsule decelerated at landing \\
    \bottomrule
    \end{tabular*}
\end{table*}

\section{Setups for the experiment}\label{sec:setup}
\subsection{Arrangement in the inner capsule}\label{subsec:inncap}
\begin{figure}[h]
    \centering
    \includegraphics[width=0.45\textwidth]{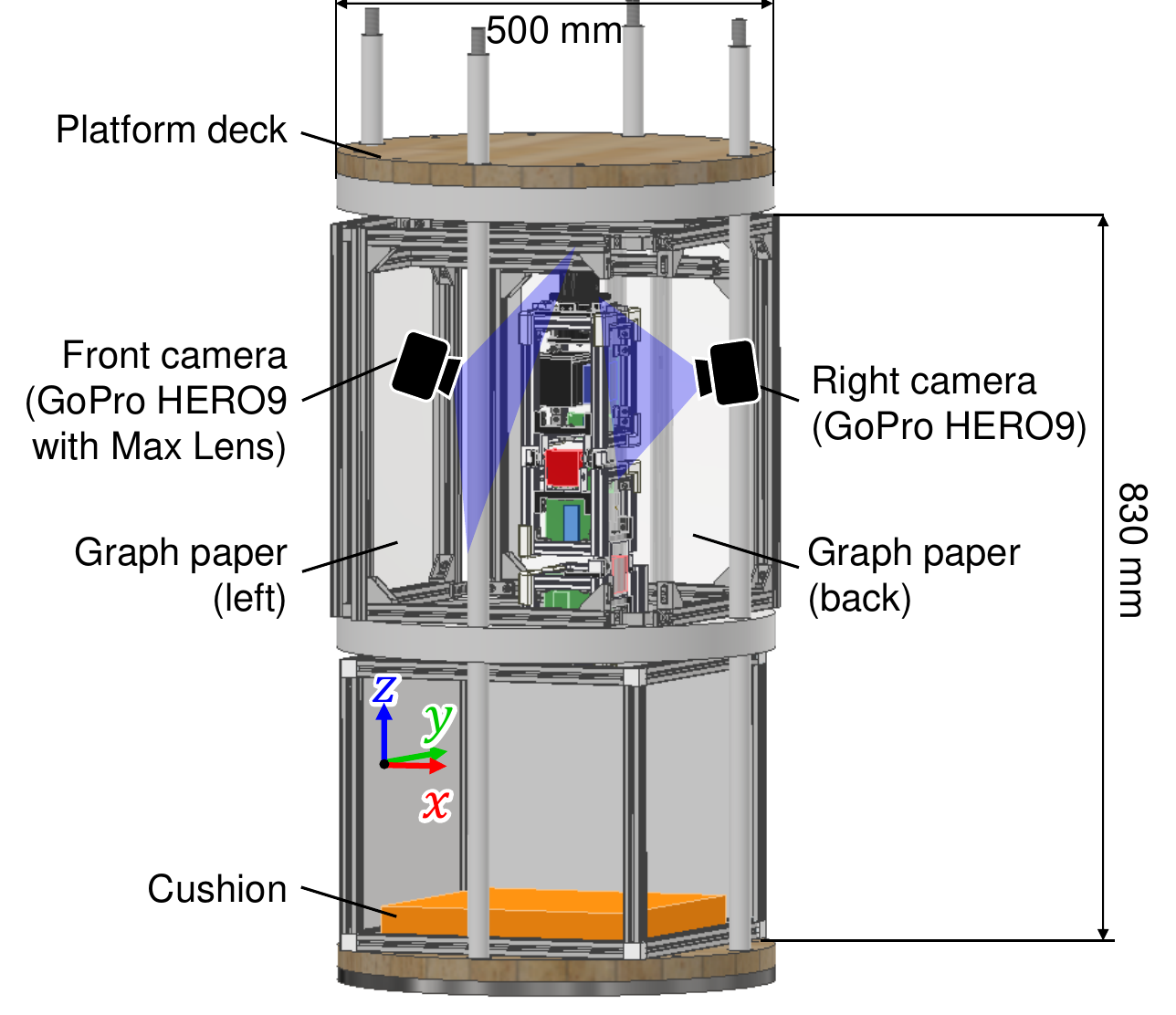} 
    \caption{Configuration of the inner capsule}\label{fig:inn}
\end{figure}

Figure \ref{fig:inn} shows the arrangement in the inner capsule. All the equipment\Del{s are} \Add{is} mounted within the envelope of the inner capsule. The test area is about 830 mm in height and 500 mm in diameter, providing sufficient space for the robot to move without touching any structures. Two action cameras (GoPro HERO9) are mounted vertically and capture the robot's motion during the sequence. Two graph papers are attached to the side walls that work as the projection screen of lasers mounted on \Del{the }body 2. The trajectories of the laser \Add{rays}\Del{lays} are captured by two additional action cameras that are used to estimate the vertical releasing velocity of the robot. A cushion is placed at the bottom of the test space, which reduces damage to the robot at landing. Other electric components to control the experiment sequence are placed on the wooden platform deck above the test space.

\subsection{Calibrations}\label{subsec:calib}
All absolute angle sensors and MEMS gyroscopes are calibrated before each test trial because every landing impact can affect their measurements. \par
\begin{figure}[h]
    \centering
    \includegraphics[width=0.4\textwidth]{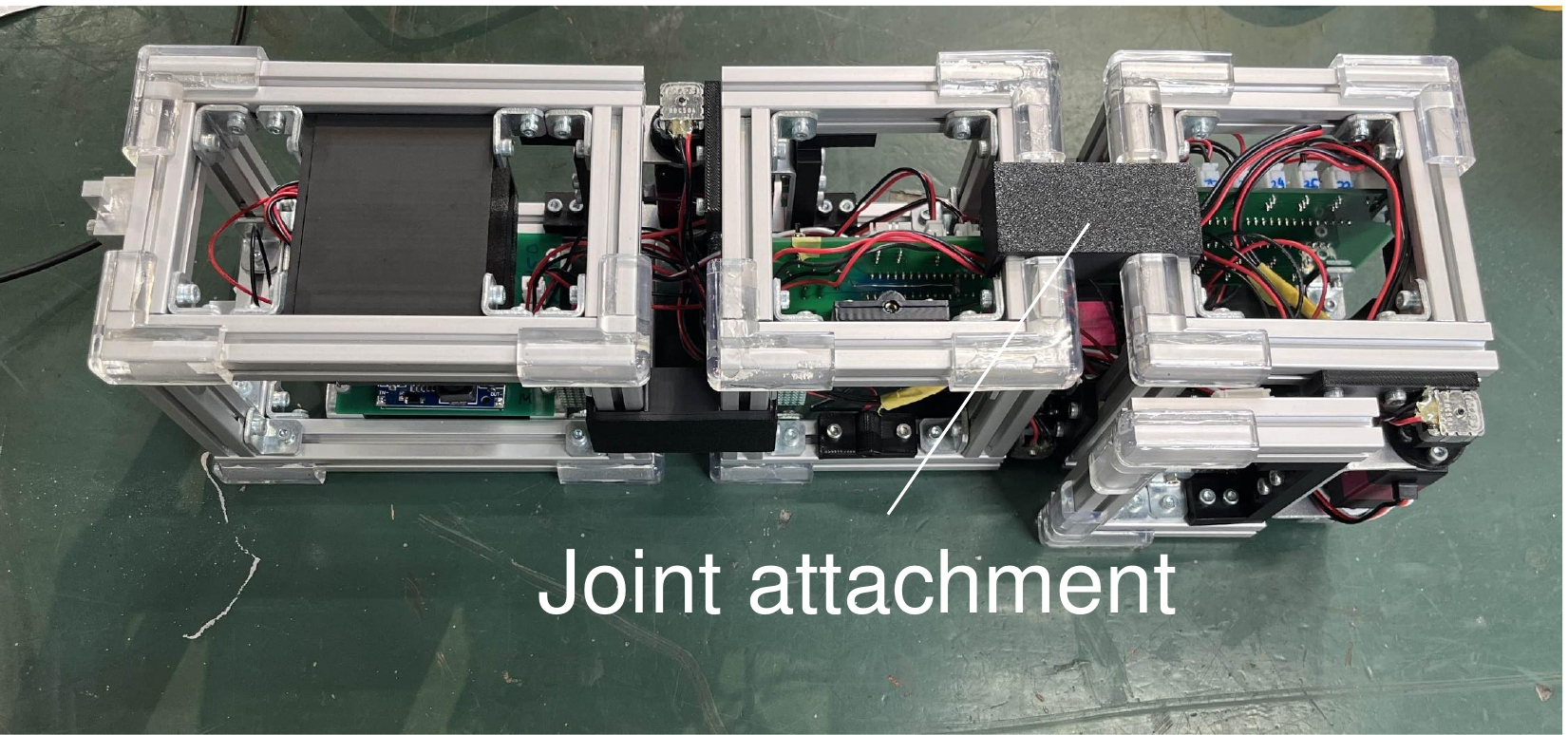} 
    \caption{Joint holding attachment for joint angle calibration}\label{fig:jattach}
\end{figure}

The absolute angle sensors are mainly used to provide a zero\Del{-} point \Add{for}\Del{of} the joint angles. The zero-point value is measured by applying an attachment that holds the frames between bodies (Fig. \ref{fig:jattach}). \par
Calibration of the MEMS gyroscope is essential to guarantee the accuracy of measurement. In general, the primary error factors of MEMS gyroscopes are zero-rate output bias, scale factor tolerance, misalignment of axes, and acceleration sensitivity. First, zero-rate output bias can be measured and trimmed a few minutes before the drop of the capsule. The duration of the test (2.5 sec) and the waiting time until a drop (a few minutes) are both sufficiently short. Therefore, the bias error is expected to have a relatively small influence compared to other factors. \par 
The scale factor is not calibrated in this experiment, and thus, $\pm3$ \% error is included in the data as shown in Table \ref{tab:mpu}. In addition, the influence of linear acceleration sensitivity is neither compensated because the gyroscope measurement is performed in microgravity and is not dependent on attitude with respect to the gravity direction.\par
\begin{figure}[h]
    \centering
    \includegraphics[width=0.45\textwidth]{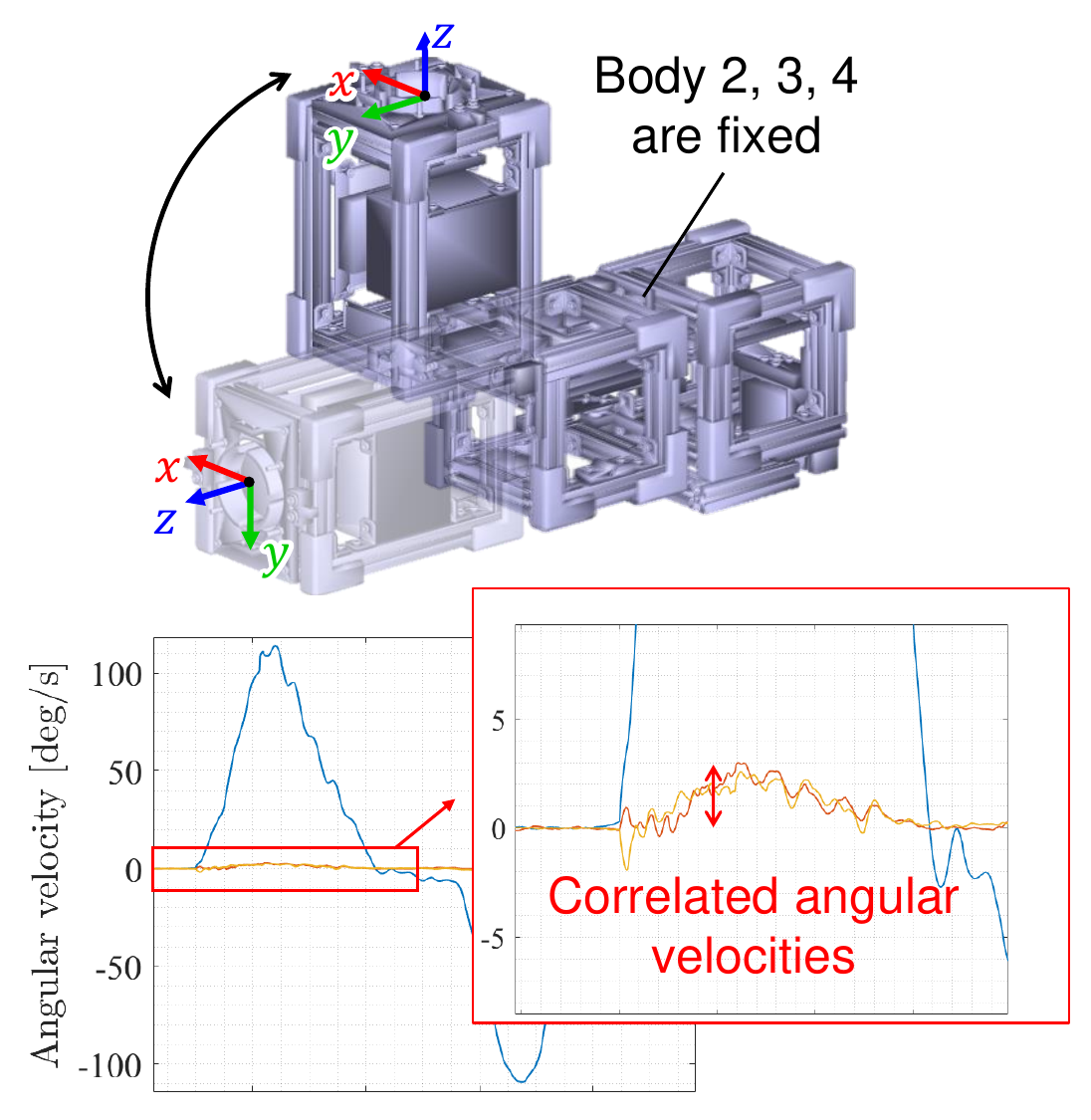} 
    \caption{Example of cross-axis correlation measurement (body 1 is rotated along joint 1)}\label{fig:crossax}
\end{figure}

 
Misalignment error is removed by measuring cross-axis correlations. Figure \ref{fig:crossax} shows an example of the cross-axis correlation measurement. In this case, body 1 is rotated along joint 1, whereas bodies 2, 3, and 4 are fixed on the ground. If the MEMS gyroscope on body 1 is ideally aligned to the rotational axis of joint 1, it exhibits the angular velocity of the $x$ component. However, the actual measurement shows correlations of $y$ and $z$ components due to misalignment. By measuring the cross-axis correlations for two orthogonal axes on each body, the actual mounted attitude of the MEMS gyroscope can be estimated. All movement patterns for the misalignment calibration are shown in Table \ref{tab:misal}. These patterns cover measurements for two orthogonal axes on each body. 
\begin{table}[h] \caption{Motion patterns for misalignment calibration}\label{tab:misal}%
    \begin{tabular}{@{}ll@{}}
    \toprule
    Rotational axis & Moved bodies \\
    \midrule
    Joint 1 & Body 1 \\
    Joint 1 & Body 2, 3 \\
    Joint 2 & Body 1, 2 \\
    Joint 2 & Body 3, 4 \\
    Joint 3 & Body 4 \\
    \bottomrule
    \end{tabular}
\end{table}

\section{Result}\label{sec:result}
This section describes the results of the drop tower test and evaluates the numerical model with the obtained data. \par
\begin{figure*}[h]
    \centering
    \includegraphics[width=0.8\textwidth]{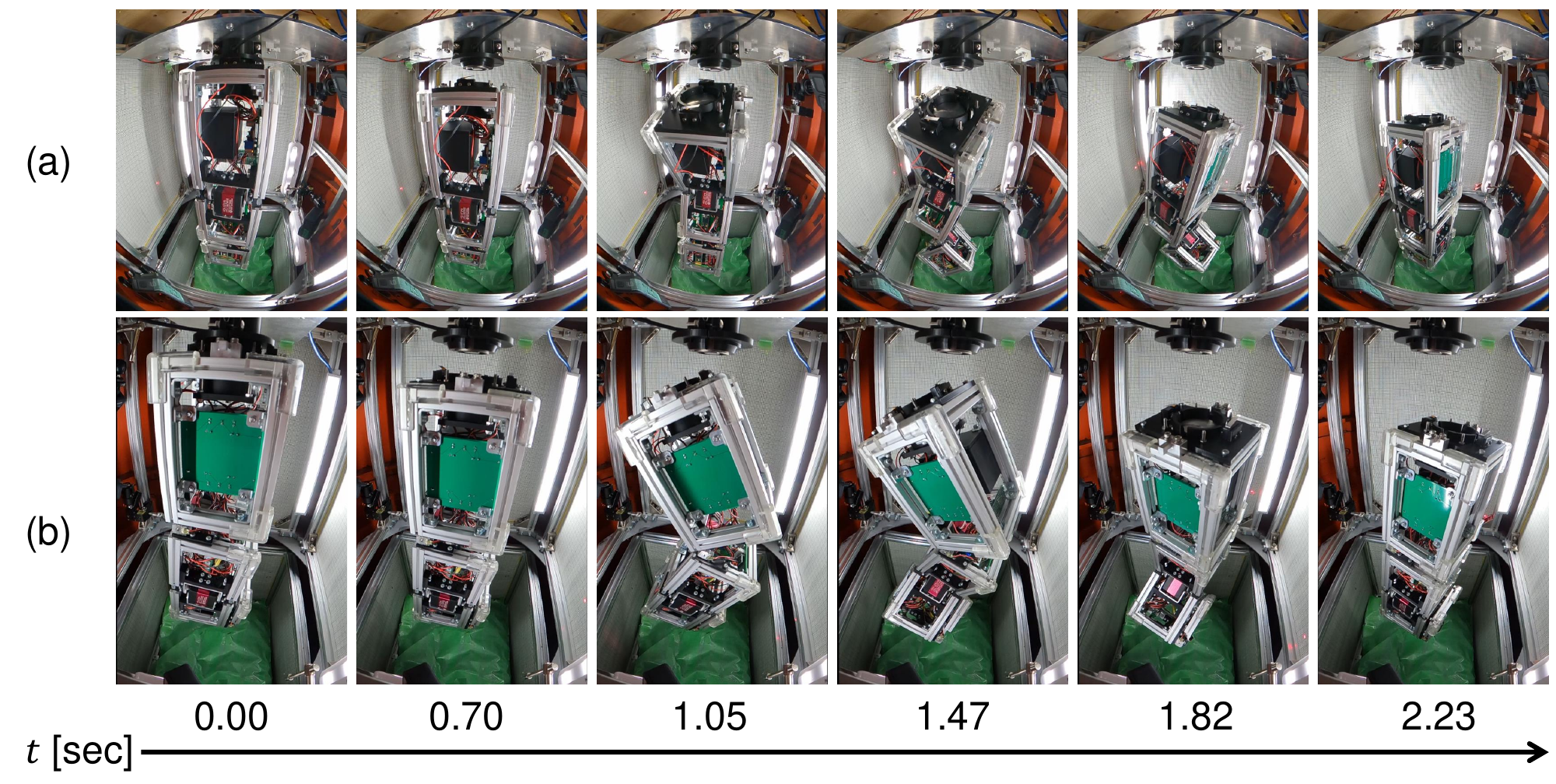} 
    \caption{Sequentially captured images (a): front camera, (b): right camera}\label{fig:frc}
\end{figure*}

Figure \ref{fig:frc} shows sequentially captured images by the front and the right cameras. The time stamps of the images are based on the sequence shown in Table \ref{tab:seq}. The figure shows that the robot's attitude is reoriented after four strokes of joint actuation as described in Section \ref{subsec:nh}. In addition, we confirmed by the video that the robot did not collide with any structures during \Add{the}\Del{an} entire maneuver. \par
\begin{figure*}[h]
    \centering
    \includegraphics[width=0.9\textwidth]{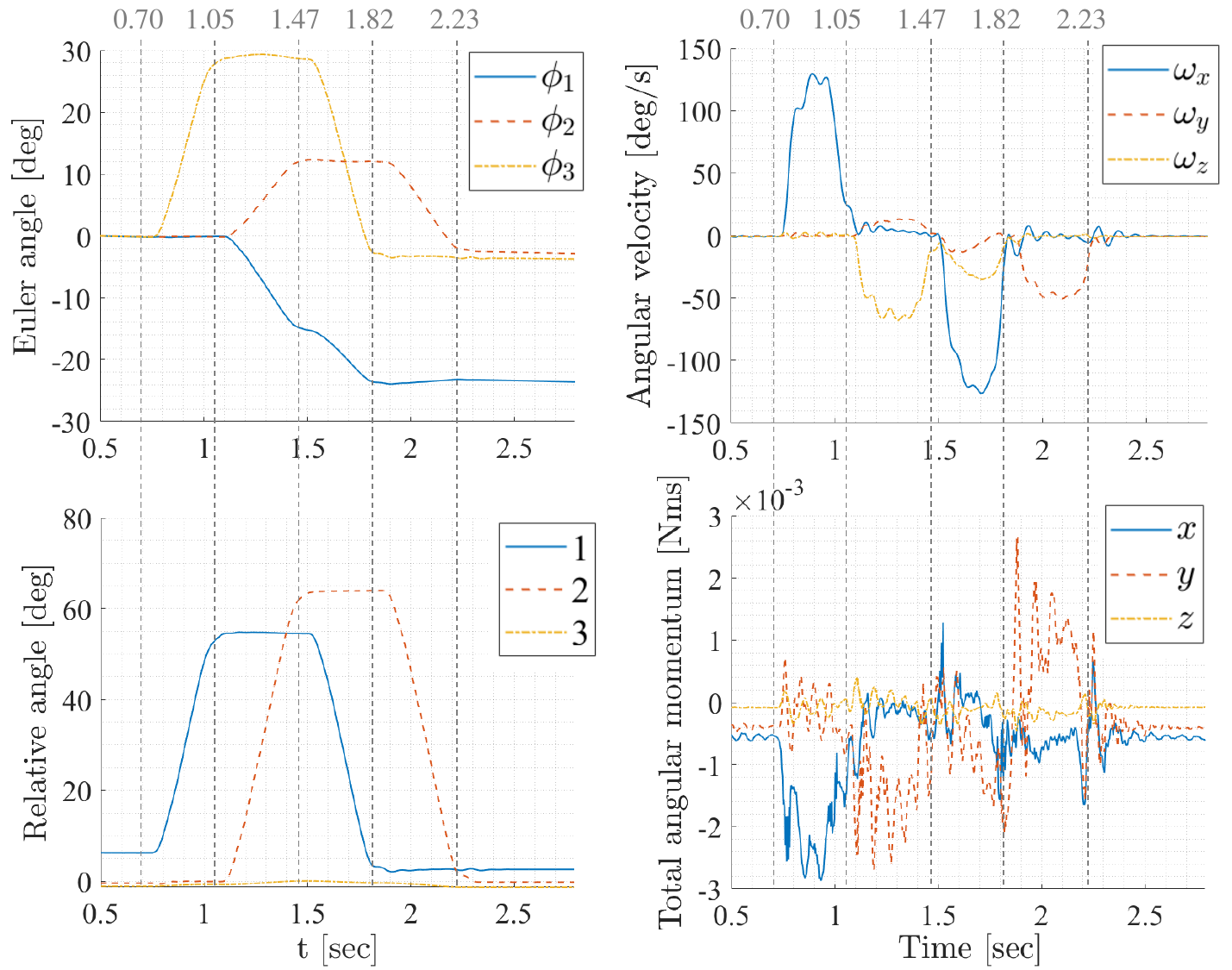} 
    \caption{Upper left: attitude of body 1 ($z$-$y$-$x$ Euler angles), upper right: angular velocity of body 1, lower left: joint angles, lower right: total angular momentum}\label{fig:euler_omg}
\end{figure*}

Data \Add{from}\Del{of} MEMS gyroscopes was successfully sampled in about $f_\mathrm{s}=280$ Hz on average. The obtained angular velocities from all bodies are processed as shown in Fig. \ref{fig:euler_omg}. The upper left figure is $z$-$y$-$x$ Euler angles of body 1, calculated by integrating the angular velocity of body 1 shown in the upper right. Initial angular velocity $\vb*{\omega_0}$ after the release was about $[\omega_{x0}, \omega_{y0}, \omega_{z0}]=[-0.73, -0.48, -0.77]$ deg/s. Total attitude reorientation after four joint strokes was 23.3 degrees, whereas the ideal reorientation angle of the designed maneuver is 26.1 degrees as described in Section \ref{subsec:opt}. This gap is generated by the initial angular velocity and the difference between the commanded joint displacements and the actual (described below).\par
The lower left figure is a history plot of joint angles. Initial angles are sampled from the absolute angle sensors, and displacements from them are calculated by integrating relative angular velocities derived by subtracting outputs of adjacent gyroscopes. Note that the initial angle of joint 1 has an offset of about $+6$ degrees due to \Add{the} gravity torque applied before a drop. In addition, joint displacements do not precisely follow the angles shown in Table \ref{tab:seq}; e.g., the actual displacement of the first stroke is 54.8 degrees, whereas the commanded angle is 59.0 degrees. This is because the reproducibility performance of our servomotor was not high. However, the obtained result is still valid because we can evaluate the numerical model by performing numerical integration with the actual joint angle displacements. \par
The lower right figure shows the total angular momentum of all bodies in the inertial frame, which is calculated with the obtained data and the numerical model described in Section \ref{subsec:model}. Ideally, the angular momentum should be conserved throughout the entire sequence, and the values before $t=0.7$ and after $t=2.5$ \Add{agree}\Del{have agreed} well. During $0.7<t<2.5$, the conservation quality gets worse due to the influence of joint actuation. The influence of linear acceleration and vibration on MEMS gyroscopes seems to induce high-frequency noise. In contrast, low-frequency noise seems to be caused by modeling errors, such as the position of the center of mass of each body.  \par
\begin{figure*}[h]
    \centering
    \includegraphics[width=0.8\textwidth]{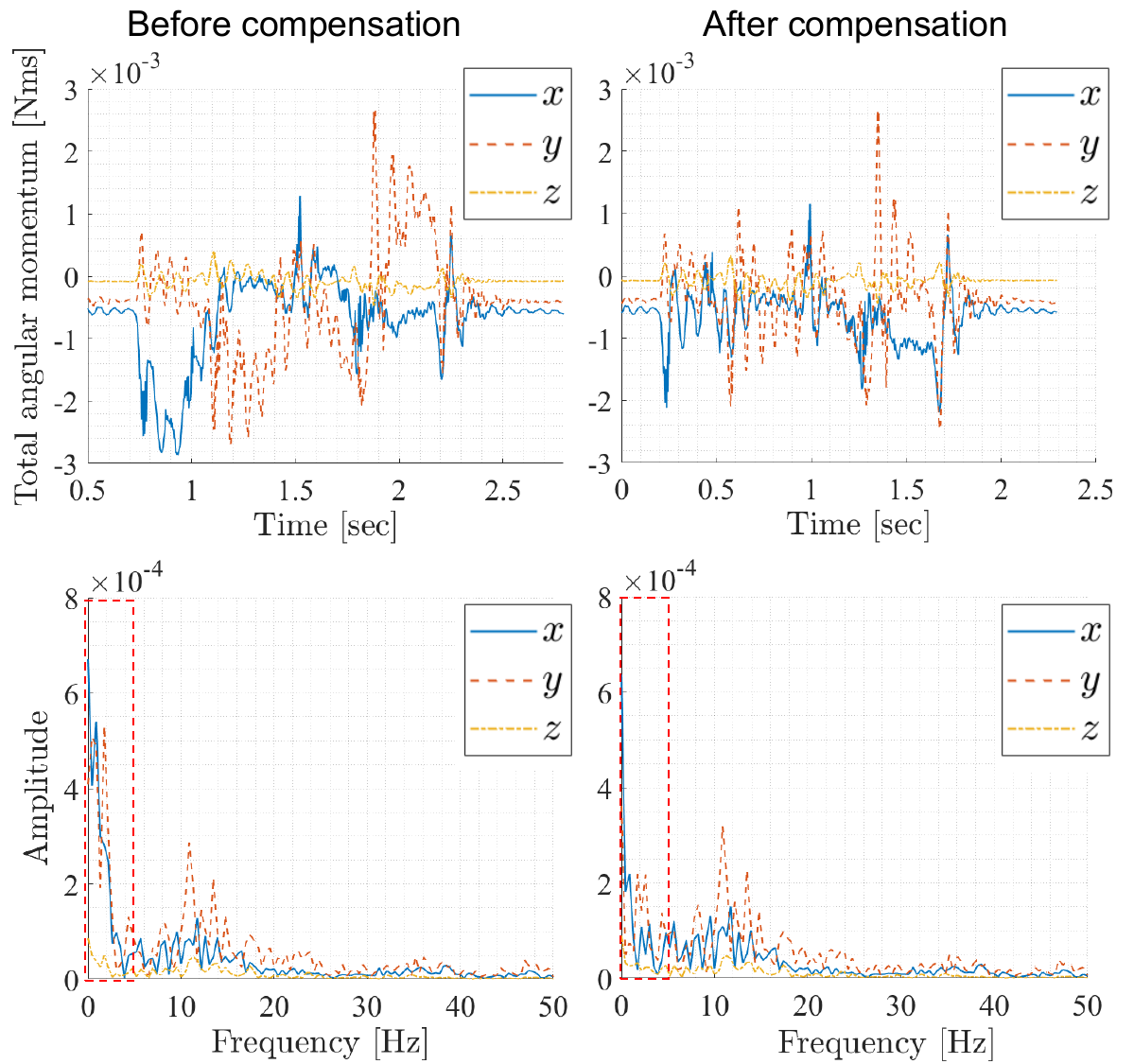} 
    \caption{Comparison of angular momentum between before and after modeling error compensation (upper: time history, lower: frequency spectrum, left: before compensation, right: after compensation)}\label{fig:amfft}
\end{figure*}

The latter error can be suppressed by compensating \Add{for} the modeling error of the center of mass of each body. Thus, we solved an optimization problem defined in Eq. \eqref{eq:cgopt} to obtain a proper offset of the center of mass. Optimized variables are $\delta\vb*{r}_{\mathrm{G}i}=[\delta x_{\mathrm{G}i}, \delta y_{\mathrm{G}i}, \delta z_{\mathrm{G}i}]$, position offsets of center of mass in the local frame of $i$-th body $(i=1,2,3,4)$. 
$\vb*{h}_{t_j}$ denotes total angular momentum of the robot at $t=t_j$ and the bar $\bar{\cdot}$ signifies moving average of local $N$-points. Thus, the objective function is the sum of the norm of the difference between \Add{the} moving-averaged angular momentum at $t=t_j$ and $t=t_0$ for all sampling points. The moving average works as a simple low-pass filter\Add{,} and \Add{the}\Del{its} width of \Add{the} averaging window $N$ is determined \Add{so that}\Del{as} it satisfies the appropriate cutoff frequency. In this case, $N=25$ is set; it corresponds to cutoff frequency $f_\mathrm{co}=5$ Hz in sampling frequency $f_\mathrm{s}=280$ Hz \citep{smith1997scientist}. 
$d_{xi},d_{yi},d_{zi}>0$ are magnitudes of maximum offsets. We set $d_{xi},d_{yi},d_{zi}=10$ mm for $i=1,2,3$ and $d_{x4},d_{y4},d_{z4}=0$ mm because the body 4 does not move with respect to the body 3 in this maneuver. Table \ref{tab:cgopt} shows the optimization result. 
\begin{mini}
    {\delta \vb*{r}_{\mathrm{G}i} (i=1\text{--}4)}          
    {J=\sum_j \qty|\bar{\vb*{h}}_{t_j}-\bar{\vb*{h}}_{t_0}|} 
    {\label{eq:cgopt}}
    {} 
    \addConstraint{-d_{xi}\leq\delta x_{\mathrm{G}i}}{\le d_{xi}} 
    \addConstraint{-d_{yi}\leq\delta y_{\mathrm{G}i}}{\le d_{yi}} 
    \addConstraint{-d_{zi}\leq\delta z_{\mathrm{G}i}}{\le d_{zi}} 
    \addConstraint{(i=1\text{--}4)}{} 
\end{mini}
\begin{table}[h] \caption{Optimal offset of the center of mass in local frames (unit: mm)}\label{tab:cgopt}%
    \begin{tabular}{@{}llll@{}}
    \toprule
    Body number & $x$ & $y$ & $z$ \\
    \midrule
    Body 1 & -2.24 &  0.92 & 2.99 \\
    Body 2 & -2.37 & -0.12 & 7.10 \\
    Body 3 & -2.76 & -1.05 & 4.92 \\
    Body 4 &  0.00 &  0.00 & 0.00 \\
    \bottomrule
    \end{tabular}
\end{table}
Figure \ref{fig:amfft} shows a comparison of angular momentum between before and after the modeling error compensation; the upper figures are time-domain, and the lower are frequency-domain \Add{spectra}\Del{spectrum} transformed by fast Fourier transformation, FFT \citep{frigo1998fftw}. The enclosed areas in the lower figure show that the compensation successfully suppressed the low-frequency noise below $f_\mathrm{co}=5$ Hz. \par
\begin{figure}[h]
    \centering
    \includegraphics[width=0.35\textwidth]{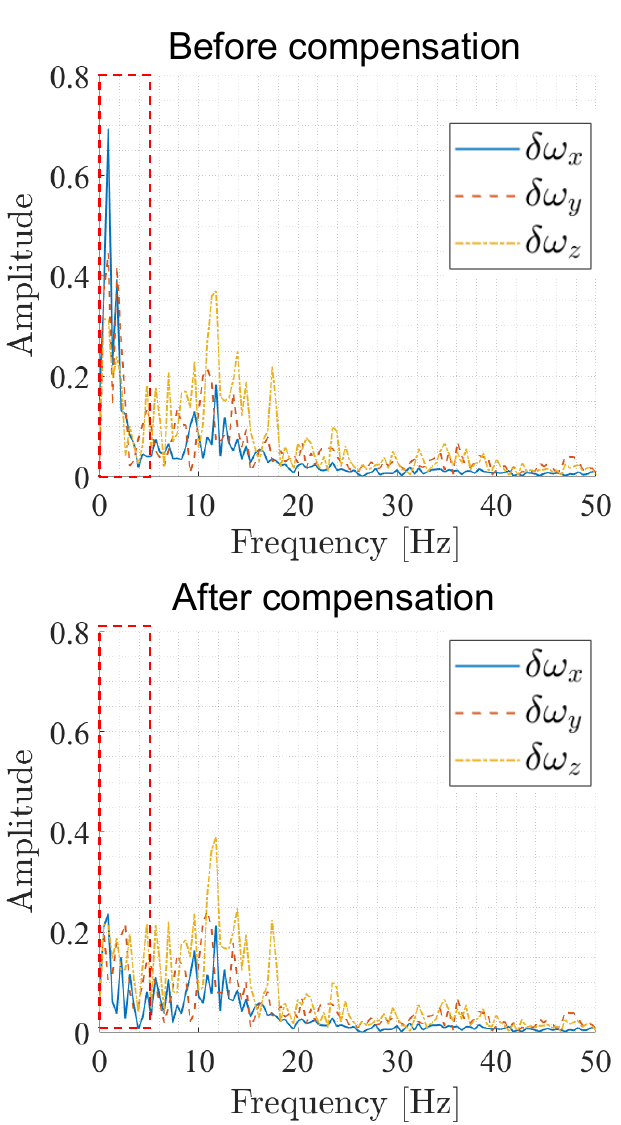} 
    \caption{Comparison of frequency spectrum of angular velocity error between before and after modeling error compensation}\label{fig:omgfft}
\end{figure}
\begin{figure}[h]
    \centering
    \includegraphics[width=0.35\textwidth]{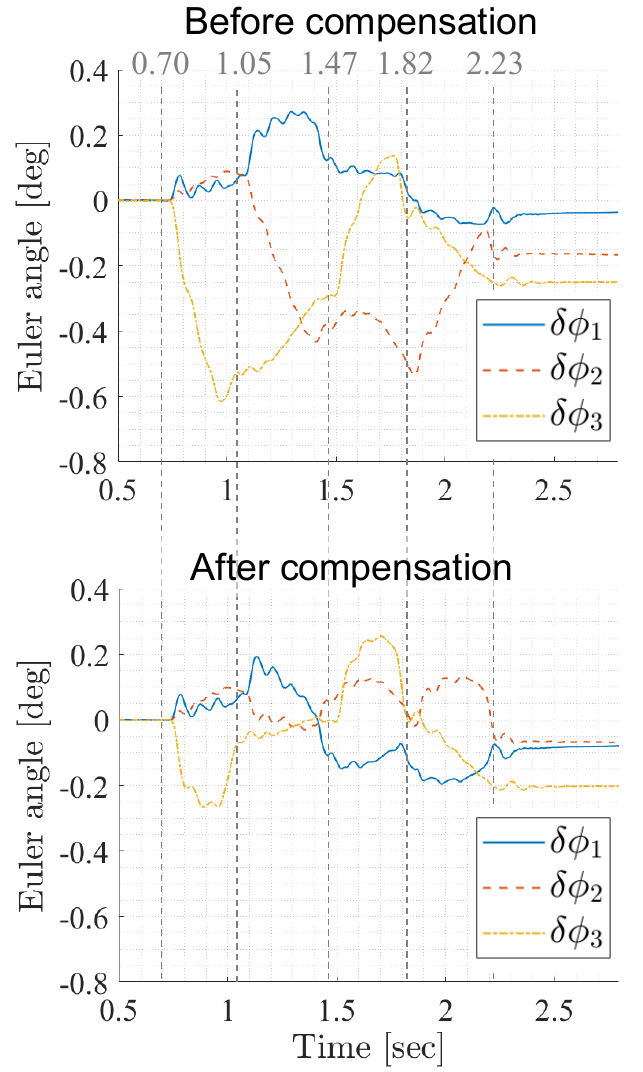} 
    \caption{Comparison of Euler angles' error between before and after modeling error compensation}\label{fig:deuler}
\end{figure}
Finally, we move on to evaluating the robot's numerical model. 
We can derive the equation of motion for the attitude of body 1 with angular momentum conservation and modeled physical properties shown in Section \ref{subsec:model}. The equation of motion is integrated with the initial attitude of body 1 and the history of joint angles. By comparing the solution of the equation of motion and the attitude directly integrated from the angular velocity of the body 1 (shown in Fig. \ref{fig:euler_omg}), we can evaluate the accuracy of the numerical model defined in Section \ref{subsec:model}.\par
Figure \ref{fig:omgfft} compares the frequency spectrum of angular velocity error between before and after modeling error compensation. As with angular momentum, the compensation successfully suppresses low\Add{-}frequency error below $f_\mathrm{co}=5$ Hz. Figure \ref{fig:deuler} compares $z$-$y$-$x$ Euler angles' error between before and after modeling error compensation. The compensation suppressed the attitude difference in the entire maneuver. \Add{Before the compensation, the final attitude error between the numerical simulation and the experiment data in $z$-$y$-$x$ Euler angle is $[\delta\phi_1,\delta\phi_2,\delta\phi_3]=[-0.03,-0.18,-0.26]$ deg, and its norm is 0.31 deg. 
On the other hand, attitude error after the compensation is $[\delta\phi_1,\delta\phi_2,\delta\phi_3]=[-0.07,-0.08,-0.21]$ deg, and its norm is 0.24 deg, which corresponds to 30 \% error reduction. The performance in this experiment is evaluated in Section \ref{subsec:orbper}.}
\Del{Thus, we concluded that the designed modeling error compensation has successfully worked.}
\par
\section{Discussion}\label{sec:discussion}
Although the hardware we used in this experiment was different from the actual flight hardware, modeling, maneuver design, and calibration methods are commonly applicable to 3U CubeSat orbit demonstrations. This section discusses the difference between the drop tower test and the orbit demonstration.
\Add{
\subsection{Expected on-orbit performance}\label{subsec:orbper} 
}
\Add{
From the results obtained in the drop tower test, we can evaluate the on-orbit performance of a spacecraft. As a potential example of the spacecraft, we consider the 3U-sized transformable CubeSat designed by the authors shown in Fig. \ref{fig:tf3u}. 
\begin{figure}[h]
    \centering
    \includegraphics[width=0.4\textwidth]{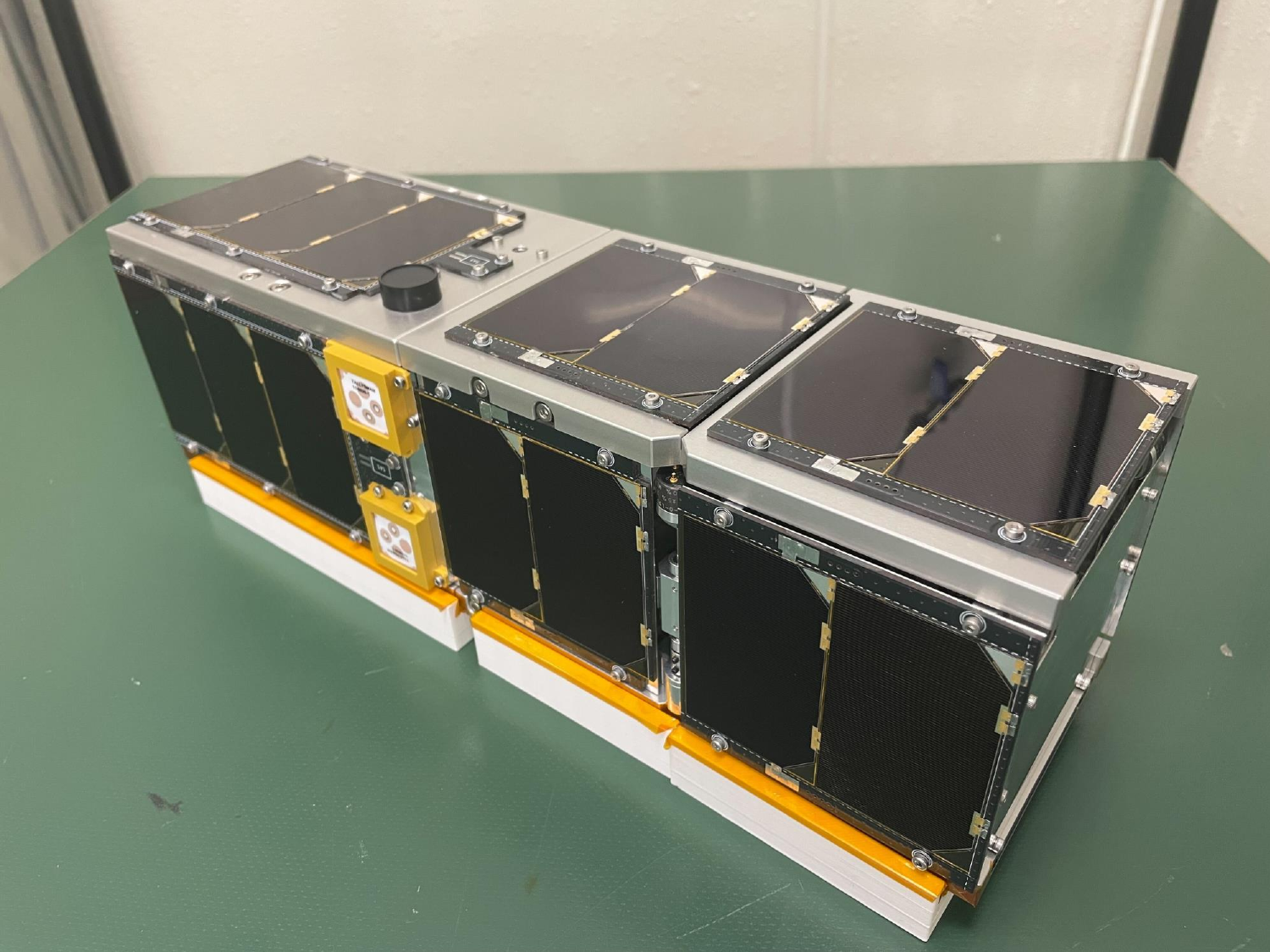} 
    \caption{\Add{Example of a 3U transformable CubeSat with similar dimensions and joint configurations to the robot used in the drop tower test}}\label{fig:tf3u}
\end{figure}
It is designed to have similar dimensions and joint configurations to the robot used in the drop tower test. However, the drop-tower robot is designed for high joint actuation speed and acceleration to complete the maneuver in the limited micro-gravity duration. In contrast, the on-orbit spacecraft actuates its joint much more slowly to maximize its torque with a smaller actuator. The joint speed performance of the drop-tower robot can be estimated by the lower left graph in Fig. \ref{fig:euler_omg} as $\dot{\theta}_\mathrm{max}=200$ deg/s. On the other hand, the designed average joint speed performance of the 3U CubeSat is $\dot{\theta}_\mathrm{ave}=30$ deg/s. 
}
\par
\Add{
In terms of reorientation maneuvers, the on-orbit CubeSat can perform various maneuvers because there is no time limit to complete the maneuver, unlike the drop tower. Figure \ref{fig:nh_dist} shows a distribution of nonholonomic attitude reorientation of the 3U transformable CubeSat where joint angles are discretized in a 5-degree interval. Joint actuation sequence is a four-stroke alternating maneuver similar to Fig. \ref{fig:ex_para} and all combinations of joint pairs are sampled to construct the distribution. The color bar shows the reorientation efficiency, which is calculated as the magnitude of the reorientation angle divided by the total joint actuation angle. The efficiency of the maneuver designed in the drop tower test was $26.11/(2\times(58.75+70.29))=0.1012$ (cf. Section \ref{subsec:opt}). The optimal maneuver in the distribution is the point shown in the yellow star in Fig. \ref{fig:nh_dist}, which corresponds to 39.31 degrees reorientation with actuation of $\Delta\theta_3=90$ deg and $\Delta\theta_2=90$ deg in this order (vice versa in the third and fourth stroke). The efficiency of the maneuver is $39.31/(2\times(90+90))=0.1092$, which is slightly better than the maneuver in the drop tower test. Thus, the expected average reorientation speed of the 3U transformable CubeSat can be estimated as $30\times 0.1092=3.28$ deg/s.
\begin{figure}[h]
    \centering
    \includegraphics[width=0.6\textwidth]{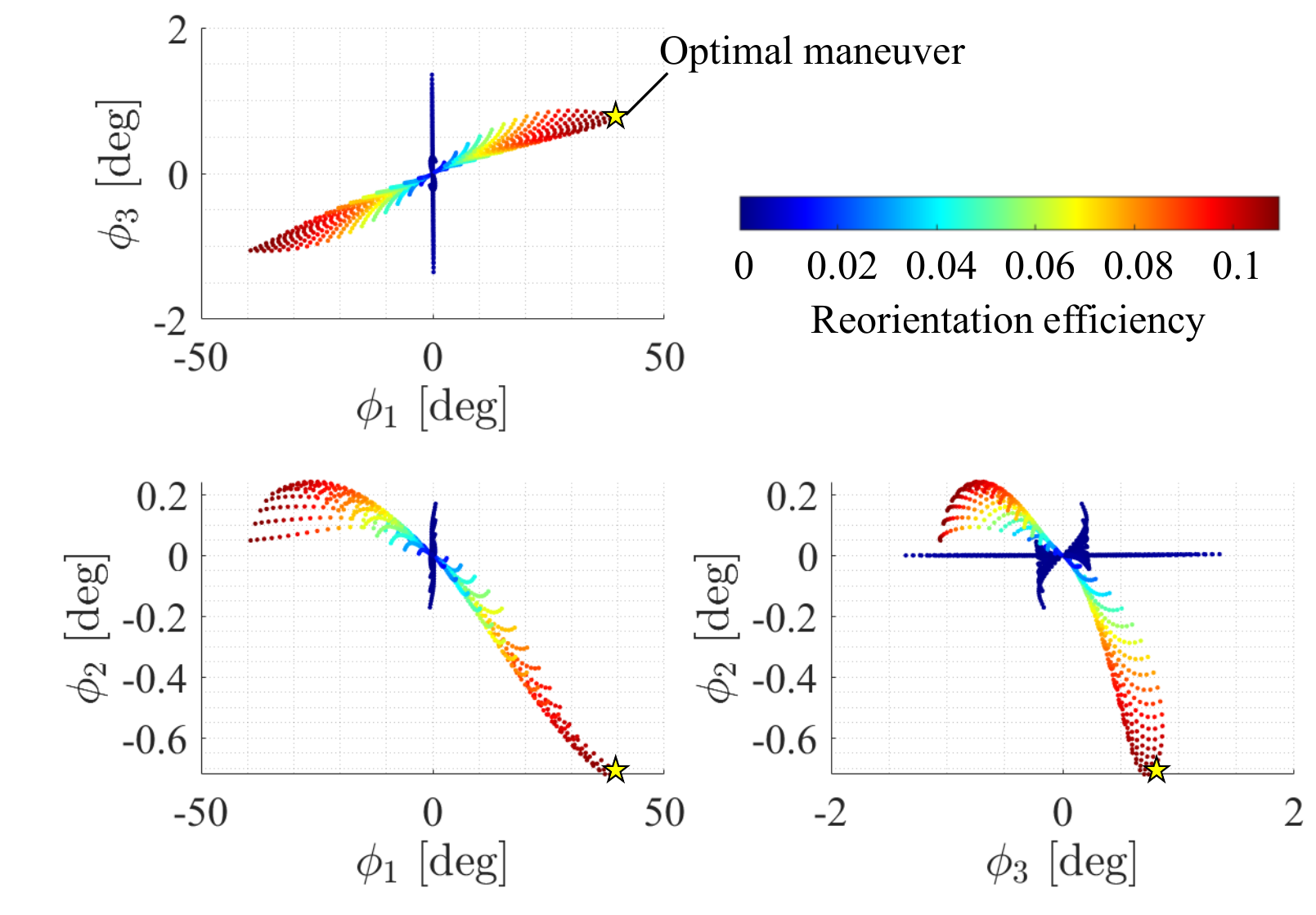} 
    \caption{\Add{Distribution of nonholonomic attitude reorientation in $z$-$y$-$x$ Euler angle} }\label{fig:nh_dist}
\end{figure}
}
\par
\Add{
Now, we can evaluate the reorientation performance of the 3U transformable CubeSat. The conclusion in Section \ref{sec:result} was that the numerical model before the modeling error compensation can design the attitude maneuver with 0.314 degree accuracy for the 23.3 degree reorientation, whereas the model after the compensation can design with 0.237 degree accuracy. When designing a 90-degree reorientation on-orbit, the accuracy is expected to be $0.314\times90/23.3=1.21$ deg with the model without the compensation, while $0.237\times90/23.3=0.92$ deg with the model with the compensation. Precise pointing cannot be achieved solely by the nonholonomic attitude control, but coarse and rapid attitude reorientation can be achieved. This accuracy seems to be the worst-case value because we can expect more precise modeling error compensation with rich on-orbit flight data as described later in Section \ref{subsec:modelerror}.
The maneuver time, assuming the average reorientation speed, is expected to be $90/3.28=27.4$ sec. 
According to previous research, the average slew rate in recent Earth observation satellite missions is around 0.1-1 deg/s, and a 3-10 deg/s slew rate is classified as a highly challenging level of agility \citep{gaude2020design}\citep{lappas2002control}. In conclusion, the drop tower test confirmed that the 3U transformable CubeSat system can overcome conventional maneuverability with acceptable accuracy. 
The reliability of this performance estimation will be examined in the future on-orbit flight experiment with the 3U transformable CubeSat.
}
\subsection{Feedback control}\label{subsec:fb} 
In the drop tower test, the robot was supposed to perform a feedforward maneuver because the duration allowed for the maneuver was strictly limited. In the orbit demonstration, however, the robot has sufficient time to perform feedback control to achieve the target attitude. A fundamental feedback control method is a successive table lookup of motion primitives \citep{ohashi_aas2018}. In each lookup, the spacecraft estimates the difference between the current and the target attitude and selects the best motion primitive to approach the target. The accuracy of the numerical model determines the accuracy of achieving the target attitude to construct the motion primitive table. Figure \ref{fig:deuler} in Section \ref{sec:result} shows that the spacecraft is expected to achieve the target attitude at least within 1 degree accuracy. 
\subsection{Modeling error compensation} \label{subsec:modelerror}
In the drop tower test, the availability of modeling error compensation is limited because the maneuver time is very short, and the robot configuration is inconsistent through trials due to structural damage experienced at every landing. In the orbit demonstration, however, more accurate compensation is expected to be possible. As well as the center of mass, the moment of inertia can be predicted by analyzing a longer series of data. A more accurate model helps the spacecraft to achieve more accurate reorientation as described in Section \ref{subsec:fb}.
\subsection{Error of gyroscopes}
We chose cost-effective MEMS gyroscopes in this experiment because many spares must be prepared due to strong landing impacts. We can expect to use \Add{a} high-performance MEMS IMU for the orbit demonstration. The comparison of the major specifications between MPU-6050 and ADIS16465-2 (an example of the high-performance IMU) is shown in Table \ref{tab:adis}. ADIS16465-2 performs much better than MPU-6050; in particular, linear acceleration sensitivity is about 10 times better. Thus, we can expect the high-frequency noise seen in Fig. \ref{fig:amfft} and \ref{fig:omgfft} to be suppressed with such high-performance IMUs.
\begin{table}[h] \caption{Specification comparison between MPU-6050 ad ADIS16465-2 (the value shown in '-' is not specified in datasheet)}\label{tab:adis}%
    \begin{tabular}{@{}lll@{}}
    \toprule
    Parameters & MPU-6050 & ADIS16465-2 \\
    \midrule
    Full-scale range & $\pm$ 500 deg/s & $\pm$ 500 deg/s \\
    Full-scale bit length & 16 bits & 32 bits \\
    Bias stability & - & 2.5 deg/hour \\
    Linear acc. sensitivity & 0.1 deg/s/g & 0.009 deg/s/g \\
    \bottomrule
    \end{tabular}
\end{table}

\section{Conclusion}\label{sec:conclusion}
The purpose of this research was to \Add{develop}\Del{construct} an evaluation method for a numerical model \Add{of}\Del{for} a transformable spacecraft\Add{,} \Add{utilizing}\Del{using} a drop tower facility. We introduced the design of the experiment and exhibited the results and analyses. 
The results showed that our numerical model replicated the robot's motion within 1 degree of accuracy. In addition, post-experiment model corrections based on the results further improved the accuracy of the numerical solution. 
The performance of the hardware we used in this experiment was limited due to cost constraints, and further studies with actual flight hardware would provide a better evaluation of the actual orbit demonstration. 

\section*{Acknowledgements}
The authors would like to thank Uematsu Electric Co., Ltd. for technical assistance with the experiments. This study was funded by a Grant for a strategic research group from the Advisory Committee for Space Engineering in Japan. 

\section*{Statements and Declarations}
\begin{itemize}
\item \textbf{Funding} This study was supported by the Advisory Committee for Space Engineering in Japan as a strategic research group. The funding supporter has not been involved in the preparation of this paper. 
\item \textbf{Competing interests} The authors have no competing interests to declare that are relevant to the content of this article.
\item \textbf{Ethics approval and consent to participate} Not applicable.
\item \textbf{Consent for publication} Not applicable.
\end{itemize}

\section*{Declaration of Generative AI and AI-assisted technologies in the writing process}
During the preparation of this work the authors used DeepL and Grammarly in order to improve readability of the text. After using this tool/service, the authors reviewed and edited the content as needed and take full responsibility for the content of the publication.


\printcredits

\bibliographystyle{model1-num-names}

\bibliography{cas-refs}





\end{document}